\documentclass[aps,pra,superscriptaddress,amssymb,showpacs,showkeys,11pt]{revtex4-1}

\usepackage{amsfonts} 
\usepackage{amsmath}
\usepackage{amssymb}
\usepackage{graphicx}
\usepackage{hyperref}
\usepackage{bm} 
\usepackage{color}
\usepackage{natbib}  
\usepackage{bm} 
\usepackage{isomath} 
\usepackage{braket}
\usepackage{subfigure}

\DeclareSymbolFont{bbgreek}{U}{bbold}{m}{n}

\providecommand{\U}[1]{\protect\rule{.1in}{.1in}}

\newcommand{\vect}[1]{\vectorsym{#1}} 
\newcommand{\ten}[1]{\tensorsym{#1}} 

\begin{document}

\title{Quantum correlations and energy currents across finite harmonic chains}

\author{Antonio A. Valido}
\email{aavalido@ull.es}
\affiliation{Instituto Universitario de Estudios Avanzados (IUdEA) and Departamento de F\'{\i}sica, Universidad de La Laguna, La Laguna 38203 Spain}

\author{Antonia Ruiz}
\affiliation{Instituto Universitario de Estudios Avanzados (IUdEA) and Departamento de F\'{\i}sica, Universidad de La Laguna, La Laguna 38203 Spain}

\author{Daniel Alonso}
\email{dalonso@ull.es}
\affiliation{Instituto Universitario de Estudios Avanzados (IUdEA) and Departamento de F\'{\i}sica, Universidad de La Laguna, La Laguna 38203 Spain}


\begin{abstract}
We present a study that addresses both the stationary properties of the energy current and quantum correlations in a three-mode chain subjected to Ohmic and super-Ohmic dissipations. An extensive numerical analysis shows that the mean value and the fluctuations of the energy current remain insensitive to the emergence of a rich variety of quantum correlations: such as, two-mode discord and entanglement, bipartite three-mode and genuine tripartite entanglement. The discussion of the numerical results is based on the derived expressions for the stationary properties in terms of the two-time correlation functions of the oscillator operators, which carry quantum correlations. Interestingly, we show that quantum discord can be enhanced by both considering initially squeezed thermal bath states and imposing temperature gradients.
\end{abstract}

\date{\today}

\maketitle
\section{Introduction}

Entanglement is one of the most striking phenomena in Quantum Physics. Composite systems exhibiting genuine quantum correlations defies our intuition, in the sense that they are not interpreted by classical or semiclassical means \cite{horodecki20091,guhne20091}. Quantum correlations are at the heart of many quantum information tasks, such as quantum teleportation and quantum communication, as well as at the core of a variety of many-body physics phenomena \cite{vedral20141}. \textsl{Quantum Thermodynamics} seeks the understanding of the emergence of thermodynamics laws from those of quantum dynamics \cite{kosloff20131,kosloff20132}. In this sense, one may wonder about the role of quantum correlations and coherence in different phenomena of interest in Quantum Thermodynamics, for instance in the thermodynamics of quantum thermal machines \cite{scarani20021,abah20121,kosloff20131,kosloff20132,hovhannisyan20131,alicki20131,correa20131,correa20141,brunner20141,gallego20141} or, more generally, in thermal non-equilibrium systems \cite{rammer20071}.

Quantum correlations have different fates depending on the environmental influence \cite{mintert20051,aolita20141}. Although entanglement is fragile with respect to thermal fluctuations and decoherence, stationary entanglement still could remain in a system subjected to dissipation \cite{wolf20111,krauter20111,ludwig20101,valido20132}. Quantum discord \cite{zurek20001,olliver20011,henderson20011} on the other hand seems more stable to environmental noise \cite{wu20111}. Recently, efforts have been made to study how dissipation may precisely drive the system onto preferred states, e.g. onto a genuine entangled state  \cite{diehl20081,kraus20081}, by engineering the interaction with environments \cite{krauter20111,manzano20131,bellomo20132,mcendoo20131}. 
 
According to the non-equilibrium theory, the analysis of the (\textsl{linear}) response of many-body systems to macroscopic thermodynamic forces, such as those induced by temperature or chemical potential gradients, and to (weak) external fields provides an opportunity to test some predictions from condensed matter theory and statistical physics. As an illustration, multipartite entanglement in spin chains has been explored through precise measurements of the magnetic susceptibility \cite{souza20081,das20131} and the heat capacity \cite{singh20131}. Also theoretical studies of these two magnitudes seem to provide observable signatures of entanglement in spin chains at thermal equilibrium \cite{wiesniak20051,wiesniak20081}. Nowadays, the energy transport through systems involving spatial continuous variables, such as chains of trapped ions, can be experimentally measured \cite{ramm20141}.

Given the increasing interest in systems under non-equilibrium thermal conditions in the quantum regime \cite{wu20111,li20121,asadian20131,prosen20141,bermudez20131,freitas20131,ruiz20141,valido20131,bellomo20131}, one may naturally ask whether the stationary response of a system to a temperature gradient may be influenced by the presence of pure quantum correlations, and in particular by \emph{genuine} multipartite entanglement. The present work tries to elucidate whether the \textit{average} properties of the stationary energy current across a harmonic chain, such the mean values and fluctuations, are sensitive to the presence of two-mode and \textit{genuine} tripartite entanglement in the system, and more generically to quantum correlations as measured by discord. Significant advances in the context of quantum spin networks indicate that the presence of bipartite entanglement does not play an important role on excitation transport \cite{plenio20081}, whereas a strong correlation between quantum coherence and transport efficiency can be present \cite{witt20131}. Although quantum correlations tend to disappear in systems subjected to a temperature gradient, it has been shown that entanglement and discord can still survive in systems under such conditions \cite{wu20111,valido20131}. Much less is known about the influence of genuine multipartite entanglement and the structure of discord on the stationary energy current in strongly dissipated harmonic chains at low temperatures. This work focuses on stationary quantum correlations in a continuous-variable system within such domain, and analyzes their possible relation to non-equilibrium conditions. 

We consider an open system model composed of a linear arrange of three harmonic oscillators, each of them interacting with its own independent heat bath. We assume that the heat baths are in an initial squeezed thermal state \cite{hu19941}. This set-up is particularly interesting in the study of the generation of entanglement between distant modes of a quantum network \cite{kuwahara20121}, and also a convenient model to analyze many issues concerning the Quantum Thermodynamics of continuous variable systems \cite{kosloff20131}. We employ the open-system formalism based on the generalized Langevin equation (GLE) \cite{hanggi20051,weiss19991,ford19882} to carry out an extensive numerical study of the stationary properties. We focus on the two- and three-mode entanglement and the discord in the presence of an energy current through the harmonic chain, for a large range of system parameters. We will analyze whether the average and the fluctuations of the energy current exhibit any evidence of the quantum correlations emerging under non-equilibrium thermal conditions.

The paper is organized as follows. In Section \ref{sec:1} we describe the model of the system and introduce the covariance matrix, which fully characterizes the stationary state of the system. Section \ref{SLA} reviews the generalized Langevin equation approach considered to obtain this state. In Section \ref{sec:3} we derive the expressions giving the average and the fluctuations of the energy current in terms of two-time correlation functions, and introduce the quatum correlations, characterized by means of the two- and three-mode entanglement and the (right) discord.  The numerical results are presented in Section \ref{SRES}, the corresponding discussion is given in Section \ref{SDIS}, and the main conclusions are put together in Section \ref{SCon}.


\section{Microscopic Model}\label{sec:1}

\begin{figure}[h]
\begin{center}
\includegraphics[scale=0.36]{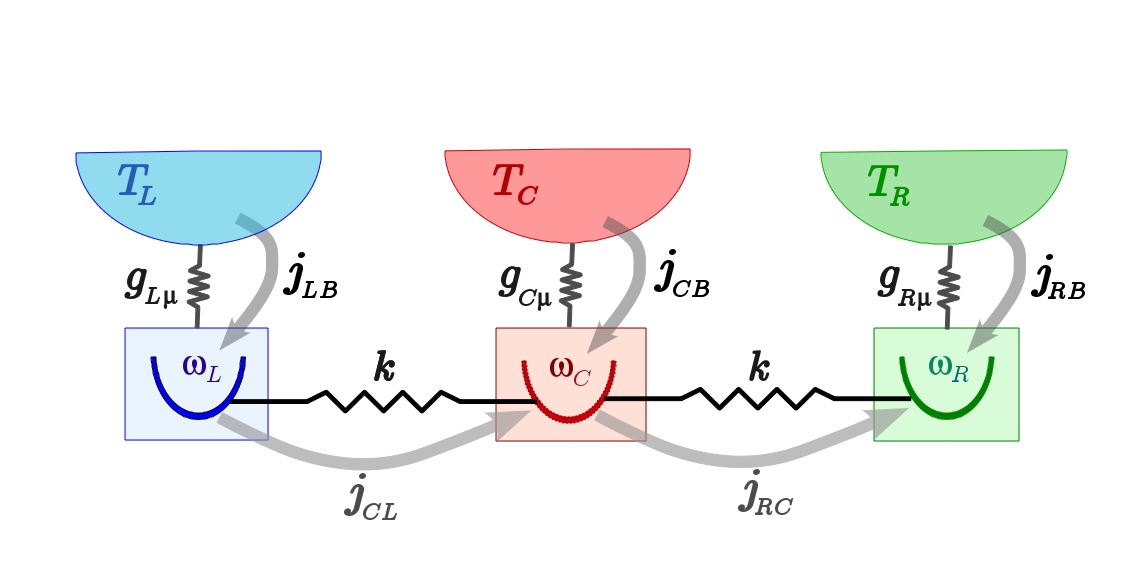}
\caption{(Color online) Schematic representation of the chain composed of the three oscillators coupled to independent heat baths, with temperatures $T_{{\cal L}}$, $T_{{\cal C}}$ and $T_{{\cal R}}$. ${\hat j}_{iB}\,(i={\cal L},\,{\cal C},\,{\cal R})$ indicates the energy current from the heat bath to the $ith$ oscillator, and ${\hat j}_{ij}$ the energy current from the $jth$ to the $ith$ oscillator, in the case of $T_{{\cal L}}>T_{{\cal C}}>T_{{\cal R}}$. $k$ is the coupling constant between first-neighbor oscillators, and $g_{i\mu}$ the coupling constant between the $ith$ chain oscillator and the $\mu th\,(\mu=1,\dots,N)$ oscillator of the bath.}\label{System}
\end{center}
\end{figure}

We consider an open one-dimensional chain composed of three harmonic oscillators, see Figure \ref{System}, labeled as ${\cal L}$ (left), ${\cal C}$ (center), and ${\cal R}$ (right),  with identical mass $m$, natural frequencies $\omega_i\,(i={\cal L},\,{\cal C},\,{\cal R})$, and position and momentum operators $({\hat x}_i,\,{\hat p}_i)$. We assume bilinear interactions between first-neighbor oscillators, ${\cal L}\leftrightarrow {\cal C}$ and ${\cal C}\leftrightarrow {\cal R}$, with strength given by a single parameter $k$. Each $ith$ oscillator is coupled with an independent heat bath composed of $N$ independent harmonic oscillators, with masses $m_{i\mu}\,(\mu=1,\dots,N)$, frequencies $\omega_{i\mu}$ and position and momentum operators $({\hat x}_{i\mu},\,{\hat p}_{i\mu})$. Eventually we will consider the  quasi-continuum limit $N\rightarrow \infty$. 
The Hamiltonian of the global system can be written as
\begin{eqnarray}
\hat H=\sum_{i={{\cal L},{\cal C},{\cal R}}}\left(\hat H_{Si}+\hat H_{Bi}\right)\,,\label{MM}
\end{eqnarray}
where
\begin{eqnarray}
\hat H_{Si}&=&\frac{\hat p^2_{i}}{2m}+\frac{1}{2}m\omega^{2}_{i}\hat x_{i}^2+\underbrace{\sum_{j={\cal L},{\cal C},{\cal R}}U_{ij}\hat{x}_{i}\hat x_{j}}_{\hat H_{I}}\,,\label{hamiltonian} 
\end{eqnarray}
with
\begin{equation}
\ten U=\frac{1}{2}\left(\begin{array}{ccc}
 k & -k & 0 \\
-k & 2k & -k\\
 0 & -k & k
\end{array}\right)\,,
\nonumber
\end{equation}
corresponds to the isolated chain, and
\begin{eqnarray}
\hat H_{Bi}&=&\sum_{\mu=1}^N  \frac{\hat{p}_{i\mu}^{2}}{2m_{i\mu}}+\frac{1}{2}m_{i\mu}\omega_{i\mu}^{2}\left(\hat{q}_{i\mu}-\frac{g_{i\mu}}{m_{i\mu}\omega_{i\mu}^{2}}\hat x_{i}\right)^2\,\label{HEI}
\end{eqnarray}
describes the three independent baths and their interactions with the oscillators, which are assumed bilinear with coupling constants $g_{i\mu}$.
The interaction term in the microscopic model given by $\hat H_{Bi}$ includes the renormalization terms
\begin{equation}
m\Delta\Omega_{i}=\sum_{\mu=1}^N\frac{g^2_{i\mu}}{m_{i\mu}\omega^2_{i\mu}}\,,
\label{}
\end{equation}
which ensures that the frequency $\omega_{i}$ is maintained as the bare frequency of the $ith$ oscillator \cite{weiss19991}, and the complete positivity of the total  Hamiltonian (\ref{MM}).

In general, a system under the influence of dissipative effects will evolve in the long time limit toward a stationary state in which any trace of its initial state has been wiped out. The initial condition is only relevant in determining the transient dynamics previous to this asymptotic state. We fix the initial state at $t_{0}\rightarrow -\infty$, and assume a barely chance of interaction between the system and the environment at to this point. Then it is reasonable to consider that the system and the environment are initially uncorrelated. As our analysis is based on quantum properties in the asymptotic stationary state, without loss of generality we will assume an initial product state given by $\hat{\rho}_{0}=\hat{\rho}_{S}\otimes(\hat{\rho}_{B{\cal L}}\otimes\hat{\rho}_{B{\cal C}}\otimes\hat{\rho}_{B{\cal R}})$ \cite{weiss19991,valido20131}, where $\hat{\rho}_{S}$ is the initial state of the isolated chain and $\hat{\rho}_{Bi}\,(i={\cal L},{\cal C},{\cal R})$ initial Gaussian quantum states corresponding to the baths, which are not necessarily at thermal equilibrium states. Assuming that initially the baths are in squeezed thermal states with zero first-moments \cite{breuer20021}, the following averages over the initial state $\hat{\rho}_{0}$ are satisfied
\begin{eqnarray}
\frac{1}{2}\left\langle\left\lbrace  {\hat q}_{i\nu}(t_{0}),{\hat q}_{i\mu}(t_{0}) \right\rbrace\right\rangle_{\hat{\rho}_{0}} &=&\delta_{\nu\mu}\,\frac{\hbar}{2m_{i\mu}\omega_{i\mu}}\left[\,1+2\,N\left(\omega_{i\mu }\right)\,+\,2\,\text{Re}\left[\,M\left(\omega_{i\mu }\right)\,\right]\,\right],  \nonumber \\
\frac{1}{2}\left\langle \left\lbrace  {\hat p}_{i\nu}(t_{0}), {\hat p}_{i\mu}(t_{0})\right\rbrace\right\rangle_{\hat{\rho}_{0}} &=&\delta_{\nu\mu}\,\frac{\hbar m_{i\mu}\omega_{i\mu}}{2}\left[\,1+2\,N\left(\omega_{i\mu}\right)\,-\,2\,\text{Re}\left[\,M\left(\omega_{i\mu}\right)\,\right]\,\right]\,,\nonumber \\
\frac{1}{2}\left\langle \left\lbrace  {\hat q}_{i\nu}(t_{0}), {\hat p}_{i\mu}(t_{0})\right\rbrace\right\rangle_{\hat{\rho}_{0}} &=&\delta_{\nu\mu}\,\hbar\,\text{Im}\left[\,M\left(\omega_{i\mu }\right)\,\right],
\label{CMHB}
\end{eqnarray}
where $\text{Re}[\,\bullet\,]$ and $\text{Im}[\,\bullet\,]$ denotes the real and imaginary part of $\bullet$, and 
\begin{eqnarray}
M_{i}(\omega_{i\mu })&=&-\cosh r_{i} \sinh r_{i}\,e^{i \theta_{i}}\,\left(\,2N_{th}\left(\omega_{i\mu }\right)\,+\,1\right), \nonumber \\
N_{i}(\omega_{i\mu })&=&N_{th}\left(\omega_{i\mu }\right)\,\left(\cosh^2 r_{i}+\sinh^2 r_{i}\right)\,+\,\sinh^2 r_{i}\,,
\nonumber
\end{eqnarray}
which satisfy the relation $|M_{i}(\omega_{i\mu })|^{2}\leq N_{i}(\omega_{i\mu })(N_{i}(\omega_{i\mu })+1)$. We have considered the same squeeze $r_i$ for all the oscillators of the $ith$ bath. $\theta_i\,(-\pi < \theta_i \leq \pi)$ is a global arbitrary rotation of the bath state $\hat{\rho}_{Bi}$, and $N_{th}(\omega_{i\mu })$ is the average occupation number of the $\mu th$ oscillator in the $ith$ bath in a thermal equilibrium state.

To induce an energy current across the chain, see Figure \ref{System}, we fix the left and right heat baths at different temperatures, $T_{{\cal L}}=T+\delta T$ and $T_{{\cal R}}=T-\delta T$ respectively, with $T$ low enough to ensure that the system remains within the quantum regime. Then we modify the temperature of the central bath, $T_{{\cal C}}=T+\Delta T$, by considering different values of $\Delta T$. This setup is particularly interesting as it makes possible to establish a quasiclassical regime in the central oscillator while maintaining the lateral oscillators in the quantum regime. Below we will show that the asymptotic stationary state derived from this manipulation of the central oscillator can exhibit a rich variety of quantum correlations, such as two-mode and bipartite three-mode and genuine tripartite entanglement.

Since the total Hamiltonian (\ref{MM}) is quadratic in both positions and momenta, and we have considered Gaussian initial bath states, the asymptotic stationary state, denoted by $\hat{\rho}_{{\cal L}{\cal C}{\cal R}}$, will be Gaussian for any initial state of the oscillators \cite{grabert19841,haake19851}. Then, the stationary quantum properties will be determined by just the first and second moments of the positions $\hat x_{i}$ and the momenta $\hat p_{i}$. The former can be made arbitrarily close to zero by unitary local transformations that do not affect the non-local properties such as entanglement. Whereas the second moments determining all the correlation properties required in our analysis are given in terms of the covariance matrix 
\begin{eqnarray}
\vect V= \begin{bmatrix}
\vect C_{\vect x \vect x}(t,t) & \vect C_{\vect x \vect p}(t,t) \\
\vect C_{\vect p \vect x}(t,t)  &  \vect C_{\vect p \vect p}(t,t), \\
\end{bmatrix}\,,\label{CVmatrix}
\end{eqnarray}
with $\hat{\vect x}=(\hat x_{{\cal L}},\hat x_{{\cal C}},\hat x_{{\cal R}})$ and $\hat{\vect p}=(\hat p_{{\cal L}},\hat p_{{\cal C}},\hat p_{{\cal R}})$, and the two-point (symmetrical) correlation functions
\begin{equation}
C_{ab}(t,t')\,=\,\frac{1}{2}\,\text{Tr}\left(\, \hat{\rho}_{0}\left\lbrace  \hat a(t), \hat b(t')\right\rbrace \,\right)\,. \label{CV}\\
\end{equation}
The second moments of the energy currents also involve the imaginary part of the two-point correlation  $\text{Tr}\left[\hat{\rho}_{0}\,\hat a(t)\,\hat b(t')\right]$, given by
\begin{equation}
Y_{ab}(t,t')\,=\,\frac{1}{2}\,\text{Tr}\left(\, \hat{\rho}_{0}\left[  \hat a(t), \hat b(t')\right] \,\right)\,. \label{CV2}\\
\end{equation}

The covariance matrix of the state $\hat\rho_{ij}$ corresponding to the subsystem defined by the $ith$ and $jth$ oscillators can be obtained from (\ref{CVmatrix}) by just taking the elements associated with these two oscillators.

It should be emphasized that at stationary conditions the correlation functions (\ref{CV}) and (\ref{CV2}) only depend on the time difference $\tau=t-t'$, and a particular initial time $t$ is irrelevant to obtain them. This will be important in what follows, when computing these stationary correlations by using the generalized Langevin equation approach.


\section{Langevin approach} \label{SLA}

Within the Langevin approach, the equations of motion that govern the evolution of the stationary correlations are derived from the microscopic model (\ref{MM}) by writing the Heisenberg equations for the oscillator positions and tracing out the degrees of freedom of the heat baths. This leads to the so-called generalized Langevin equation,
\begin{equation}
m\,\ddot{\hat x}_{i}+m\Omega_{i}^2\,\hat x_{i}+U_{ij}\,\hat x_{j}-\frac{1}{\hbar}\int_{t_{0}}^{t}d\tau\,\chi_{i}(t-\tau)\,\hat x_{i}(\tau)=\hat F_{i}(t)\,,
\label{LEq}
\end{equation}
where we have introduced the potential
\begin{eqnarray}
\Omega_{i}^2\,=\,\omega_{i}^2 +\Delta\Omega_{i}=\omega_{i}^2+\frac{1}{\pi m}\int_0^\infty d\omega ~\frac{J_i(\omega)}{\omega}\,,
\end{eqnarray}
the susceptibilities
\begin{eqnarray}
\chi_{i}(t)\,=\,\frac{2\hbar}{\pi}\,\Theta(t)\int_{0}^{\infty}d\omega J_{i}(\omega)\,\sin\left(\omega t\right)\label{sus}
\end{eqnarray}
and the fluctuating forces
\begin{eqnarray}
\hat F_{i}(t)\,=\,\sum_{\mu=1}^{N}g_{i\mu}\Big(\hat x_{i\mu}(t_{0})\cos\left(\omega_{i\mu}(t-t_{0})\right)+\frac{\hat p_{i\mu}(t_{0})}{m_{i\mu}\omega_{i\mu}}\sin\left(\omega_{i\mu}(t-t_{0})\right)\Big),  \label{FF} 
\end{eqnarray}
with $\Theta(t)$ the Heaviside step function, and the spectral density of the environment given by 
\begin{eqnarray}
J_{i}(\omega)=\frac{\pi}{2}\sum_{\mu=1}^{N}\frac{g^2_{i\mu}}{m_{i\mu}\omega_{i\mu}}\,\delta\left(\omega-\omega_{i\mu}\right)\,.\label{Jden}
\end{eqnarray}
As long as the stationary solution of the generalized Langevin equation is guaranteed, one may take the limit $t_{0}\rightarrow -\infty$ in Eq. (\ref{LEq}), and then, use the Fourier transform $\tilde{x}_{i}(\omega)=\int dt e^{i\omega t}\hat x_{i}(t)$ to obtain the stationary solution of the position and momentum operators \cite{ford20011,ludwig20101,valido20131}. By replacing these solutions into the correlation elements (\ref{CV}) and averaging over the initial state $\hat{\rho}_{0}$, it follows
\begin{eqnarray}
\left(
\begin{array}{c}
C_{x_{i}x_{j}}(t,t') \\
C_{x_{i}p_{j}}(t,t') \\
C_{p_{i}p_{j}}(t,t') \\
\end{array}
\right)\,=\,
\frac{\hbar}{2}\int\frac{d\omega}{2\pi}\int\frac{d\omega'}{2\pi}\,e^{-i(\omega t-\omega' t')}
\left(
\begin{array}{c}
1 \\
im\omega' \\
m^2\omega\omega' \\
\end{array}
\right)
G_{ij}(\omega,\omega')\,,\label{SLEq1}
\end{eqnarray}
with
\begin{equation}
G_{ij}(\omega,\omega')=\sum_{l,m={\cal L},{\cal C},{\cal R}}\tilde{ \alpha}_{il}(\omega)\left\langle \left\lbrace \tilde{F}_{l}(\omega), \tilde{F}_{m}(-\omega')    \right\rbrace  \right\rangle_{\hat{\rho}_{0}} \tilde{ \alpha}_{mj}(-\omega')\,\label{Gij}
\end{equation}
and
\begin{equation}
\tilde{\ten{\alpha}}(\omega)=\left(\ten {\Gamma}+\ten U\right)^{-1}, 
\label{Greenf}
\end{equation}
where ${\Gamma}_{ij}=\delta_{ij}\Big(-m\omega^2 +m\Omega_{i}^2-\frac{1}{\hbar}\tilde{\chi}_{i}(\omega)\Big)$.

The two-points correlation functions $Y_{ab}(t,t')$ (\ref{CV2}) satisfy an expression identical to (\ref{SLEq1}), but replacing in $G_{ij}(\omega,\omega')$ the anticommutator of the fluctuating force by the commutator.

Expression (\ref{Greenf}) is nothing but the Fourier transform of the (matrix) Green function \cite{breuer20021,dhar20071} for the generalized Langevin equation (\ref{LEq}). The real and imaginary parts of the Fourier transform $\tilde{\chi}_{i}(\omega)$ of the susceptibility (\ref{sus}) are  given by (see Appendix \ref{ASusc} for further details)
\begin{eqnarray}
\text{Im}\left[\,\tilde{\chi}_{i}(\omega)\,\right]&=&\hbar\left[\,\Theta(\omega)\,J_{i}(\omega)-\Theta(-\omega)\,J_{i}(-\omega)\,\right], \label{ImX} \\
\text{Re}\left[\,\tilde{\chi}_{i}(\omega)\,\right]&=&\frac{1}{\pi}\,P\int\frac{\text{Im}\left[\,\tilde{\chi}_{i}(\omega')\,\right]}{\omega'-\omega}\,d\omega' , 
\label{ReX}
\end{eqnarray}
where $P$ denotes the Cauchy principal value. The second expression is the well-known Kramers-Kronig relation \cite{weiss19991} arising from the causal nature of the susceptibility. 

In order the three-mode system can reach a stationary state, the function $\alpha_{ij}(t)$ must approach a combination of decaying exponentials in the long time limit \cite{jung19851}. According to a previous study of the equations of motion of the system-plus-environment complex in terms of normal modes \cite{dhar20071}, the existence of a well defined stationary solution entails that $\left( \tilde{\alpha} _{ij}(\omega)\right)^{-1}$ has no any real root $\Omega_{b}$ corresponding to the frequency of a bound normal mode, which implies that $\text{Im}\left[\,\tilde{\chi}_{i}(\Omega_{b})\,\right]\neq 0$. From Eq. (\ref{ImX}), the latter condition means that $\Omega_{b}$ must be contained within the domain of the bath spectral density $J_{i}(\omega)$ \cite{dhar20071}. In general, the heat baths can be considered as composed of a large number of degrees of freedom with finite broad band spectrum, in which the most energetic environmental-degree is roughly determined by a cut-off frequency $\omega_{c}$. This ensures that the natural frequencies of the system are well embedded in the environmental spectrum, and consequently it makes possible an irreversible energy transfer from the system to the environment, at least in a finite time much larger than the natural time scale of the system. We shall impose $\omega_{c}>>\sqrt{\omega^2_{i}+ k/m}\,\,(i={\cal L},{\cal C},{\cal R})$ in order to ensure an irreversible evolution of the three-oscillator chain toward a well defined asymptotic stationary state.  

The covariance matrix of this stationary state is completely determined by the correlation functions (\ref{SLEq1}) evaluated at equal time, once the correlation functions of the fluctuating forces (\ref{FF}) have been obtained. These correlations depend only on the initial environmental state. Below we show the relation between the fluctuating forces and the initial covariance matrix (\ref{CMHB}) of the heat baths.


\subsection{Fluctuation-Dissipation Relation}


Our choice of the initial environmental state implies the statistical independence of the fluctuating forces corresponding to different heat baths, \emph{i.e.} $\left\langle\,\left\lbrace {\hat F}_{l}(t),{\hat F}_{m}(t')\right\rbrace\,\right\rangle=0$ for all $l\neq m$. Whereas, according to Eqs. 
(\ref{CMHB}), the symmetrical two-time correlation function of the fluctuating forces associated with a given $lth$ bath is given by (see Appendix \ref{AFF} for further details)
\begin{eqnarray}
&&\frac{1}{2}\left\langle \left\lbrace \hat{F}_{l}(t),\hat F_{l}(t')\right\rbrace\right\rangle_{\hat{\rho}_{0}}\,=\,\sum_{\mu=1}^N\frac{\hbar\,g_{l\mu }^2}{m_{l\mu} \,\omega_{l\mu }}\Bigg[\left(\frac{1}{2}+N_{l}(\omega_{l\mu})\right)\cos\left(\omega_{l\mu}(t'-t)\right) \nonumber \\
&&+\text{Re}\left[\,M_{l}(\omega_{l\mu})\,\right]\cos\left(\omega_{l\mu}(t+t'-2t_{0})\right)+\text{Im}\left[\,M_{l}(\omega_{l\mu})\,\right]\sin\left(\omega_{l\mu}(t+t'-2t_{0})\right)\Bigg]\,.
\label{FFE}
\end{eqnarray}
The average of the corresponding commutator can be expressed as
\begin{eqnarray}
\dfrac{1}{2}\left\langle\left[{\hat F}_{l}(t),{\hat F}_{m}(t')\right]\right\rangle_{\hat{\rho}_{0}}=
i\,\delta_{lm}\,\frac{\hbar}{2}\sum_{\mu=1}^N \frac{g_{l\mu}^2}{m_{l\mu}\omega_{l\mu}}\,\sin\left(\omega_{l\mu}(t'-t)\right)\,.
\label{AFF2a}
\end{eqnarray}
The dependence of the symmetrical two-time correlation functions on the initial time $t_0$ is eliminated in the case of an initial thermal equilibrium state of the $lth$ bath, in which $M_{l}(\omega_{l\mu})=0$ and $N_{l}(\omega_{l\mu})=N_{th}(\omega_{l\mu})$ for all $\mu$ values. Although in the previous section we have already fixed the time limit $t_{0}\rightarrow -\infty $ in order to obtain the stationary solution, we shall maintain the notation $t_{0}$ for convenience in order to make more clear the following discussion.

As shown in Appendix \ref{AFF}, the non-stationary terms in Eq.(\ref{FFE}) come from the average of factors involving $a_{i\mu}^{\dagger}a_{i\nu}^{\dagger}$ and $a_{i\mu}a_{i\nu}$, with $a_{i\mu}\,(a_{i\mu}^{\dagger})$ the annihilation (creation) operator of the $\mu th$ mode in the $i$th reservoir. These terms describe non-conservative energy processes that take place in the heat bath at the initial time $t_{0}$, and therefore, they may influence the transient dynamics of the three-oscillator chain. However, they become highly oscillatory in the long time limit $(\,(t+t'-t_{0})\rightarrow \infty )$ and their contribution to the stationary properties may be disregarded \cite{hu19941}. When taking the quasi-continuum limit $\smash{\sum_{\mu}} \rightarrow \smash{\int d\omega}$ in the environment spectral density, only the stationary term in Eq.(\ref{FFE}) remains. This assertion holds for an environment with a broad spectrum limited by $\omega_c$, and a finite interaction between the reservoir modes and the system oscillators. Mathematically, the latter translates into that the spectral densities $J_{l}(\omega)$ are finite continuous functions, and the corresponding coupling strengths should decay at least as $1/\omega^{2}$ at high frequencies. Under these conditions, the long time limit of the symmetrical two-time correlation function (\ref{FFE}) reduces to the following expression in the frequency domain (see Appendix \ref{AFF} for details)
\begin{equation}
\frac{1}{2}\left\langle \left\lbrace \tilde{F}_{l}(\omega), \tilde{F}_{l}(\omega')\right\rbrace\right\rangle_{\hat{\rho}_{0}}= 2\pi\,\delta(\omega+\omega')\,\text{Im}\left[\,\chi_{l}(\omega)\,\right]\coth\left(\frac{\hbar\,\omega}{2\,k_{B}\,T_{l}}\right)\cosh\left(2r_{l}\right)\,.
\label{CFFDAL}
\end{equation}
The average of the corresponding commutator reduces to
\begin{equation}
\frac{1}{2}\left\langle \left[ \tilde{F}_{l}(\omega), \tilde{F}_{l}(\omega')\right]\right\rangle_{\hat{\rho}_{0}}= 2\pi\,\delta(\omega+\omega')\,\text{Im}\left[\,\chi_{l}(\omega)\,\right]\,.
\label{CFFDAL2}
\end{equation}

Similar results have been previously obtained within the path integral formalism \cite{hu19941,fleming20131}, see also \cite{pagel20132}.

We point out that an initially squeezed state of the environment makes the reduced system to \textit{notice} an effective temperature above the temperature $T_{i}\,(i \in \lbrace {\cal L},{\cal C},{\cal R} \rbrace)$ of the heat bath at thermal equilibrium. This effect has interesting consequences in the efficiency of thermal machines within the quantum regimen \cite{correa20141}. 

Now we can replace the autocorrelations (\ref{CFFDAL}) into the expressions (\ref{Gij}), and perform the integral in the frequency $\omega'$ to obtain a closed-form expression for the two-time correlation functions $C_{ab}(t,t')$. Notice that $C_{ab}(t,t')=C_{ab}(\tau=t-t',0)$ due to the stationary condition of the fluctuating force correlation. A similar procedure is followed for the functions $Y_{ab}(t,t')$ (\ref{CV2}).

In general, there are not analytic expressions giving the integrals involved in the correlation functions in terms of the system parameters, such as the bath temperatures, the oscillator frequencies and the coupling strengths. We will compute them by means of numerical methods.


\section{Energy current and Quantum Correlations}\label{sec:3}


The non equilibrium conditions imposed by the different bath temperatures drive an energy current through the system, see Figure \ref{System}. A discrete definition of the energy currents associated with each chain oscillator can be derived from its local energy \cite{dhar20081,lepri20031,ruiz20141}
\begin{equation}
{\hat h}_i=\frac{\hat p^2_{i}}{2m}+\frac{1}{2}m\omega^{2}_{i}\hat x_{i}^2+{\hat u}_i({\hat {\vect x}})+\frac{1}{4}\sum_{\mu=1}^N m_{i\mu}\omega_{i\mu}^{2}\left(\frac{g_{i\mu}}{m_{i\mu}\omega_{i\mu}^{2}}\hat x_{i}-\hat{q}_{i\mu}\right)^2\,,
\end{equation}
with ${\hat u}_i({\hat {\vect x}})=({\hat x}_i-{\hat x}_{\cal C})^2/4$ for $i=({\cal L}, {\cal R})$, and ${\hat u}_{\cal C}({\hat {\vect x}})=[({\hat x}_{\cal C}-{\hat x}_{\cal L})^2+({\hat x}_{\cal C}-{\hat x}_{\cal R})^2]/4$. The time derivative of ${\hat h}_i$ leads to the discrete continuity equations
\begin{equation}
\frac{d{\hat h}_i}{dt}={\hat j}_{i}(t)+{\hat j}_{i{\cal B}}(t) 
\end{equation}
with ${\hat j}_{i}(t)={\hat j}_{i\,{\cal C}}(t)$ for $i=({\cal L}, {\cal R})$, and ${\hat j}_{\cal C}(t)={\hat j}_{{\cal C}{\cal L}}(t)+{\hat j}_{{\cal C}{\cal R}}(t)$. The term
\begin{eqnarray}
\hat{j}_{ij}(t)=-\hat{j}_{ji}(t)=\frac{k}{4m}\Big[\left\lbrace\hat x_{j}(t),\hat p_{j}(t)\right\rbrace-\left\lbrace\hat x_{i}(t),\hat p_{i}(t)\right\rbrace+\underbrace{\Big( \left\lbrace \hat x_{j}(t),\hat p_{i}(t)\right\rbrace-\left\lbrace\hat x_{i}(t),\hat p_{j}(t) \right\rbrace \Big)}_{\text{\normalsize{Correlation terms}}}\Big] \label{J1}
\end{eqnarray}
can be identified as the energy current from the $jth$ oscillator to the $ith$ oscillator, whereas
\begin{eqnarray}
{\hat j}_{i{\cal B}}(t)&=&\frac{1}{4m}\sum_{\mu=1}^N \Bigg( g_{i\mu} \Big( \left\lbrace\hat q_{i\mu}(t),\hat p_{i}(t)\right\rbrace-\frac{g_{i\mu}}{m_{i\mu}\omega_{i\mu}^{2}}\left\lbrace\hat x_{i}(t),\hat p_{i}(t)\right\rbrace-\frac{m}{m_{i\mu}}\left\lbrace\hat x_{i}(t),\hat p_{i\mu}(t)\right\rbrace\Big) \nonumber \\
&+&m\omega_{i\mu}^{2}\left\lbrace\hat q_{i\mu}(t),\hat p_{i\mu}(t)\right\rbrace\Bigg)
\end{eqnarray}
corresponds to the energy current from the $ith$ heat bath into the $ith$ oscillator. At stationary conditions the total current coming from the baths into the system becomes zero. Here we will focus on the analysis of the total current flowing from the ${\cal L}$- to the ${\cal R}$-oscillator, defined as
\begin{equation}
\hat{J}(t)=\hat{j}_{{\cal C}{\cal L}}(t)+\hat{j}_{{\cal R}{\cal C}}(t)\,.
\label{GEF1}
\end{equation}

Our study is based on the stationary properties of the total energy current, which are basically determined by its first- and second-moments, or equivalently, by its average and fluctuations. The steady state average of the total energy current 
\begin{equation}
\left\langle\hat{J}\right\rangle=\left\langle\hat{j}_{{\cal R}{\cal C}}\right\rangle+\left\langle\hat{j}_{{\cal C}{\cal L}}\right\rangle \label{TEC}
\end{equation}
can be obtained by tracing (\ref{J1}) over the the stationary state and using the stationary solutions of the two-time correlation functions (\ref{SLEq1}), which leads to
\begin{eqnarray}
\left\langle \hat{j}_{ij}\right\rangle &=& \frac{k}{2m}\left[\,C_{x_{j}p_{j}}(t,t)-C_{x_{i}p_{i}}(t,t)+\left(C_{x_{j}p_{i}}(t,t)-C_{x_{i}p_{j}}(t,t)\,\right)\,\right]\,
 \label{EF1} 
\end{eqnarray}
for the local currents $\hat{j}_{{\cal C}{\cal L}}$ and $\hat{j}_{{\cal R}{\cal C}}$. 

Since the quantum correlations shared by the oscillators, in particular entanglement, are partially encoded on the correlation terms indicated in (\ref{J1}), one might expect that the energy current could be sensitive to these correlations. Notice that the total current involves the correlations between the central and the side oscillators, while it does not depend on the crossed correlation function between the two side oscillators.  We shall further analyze this issue in Section (\ref{SDIS}).


\subsection{Fluctuations of the energy current}


To have a better understanding of the system behavior under non-equilibrium conditions we also study the two-time correlation functions of the energy currents (\ref{J1}). The fluctuations  can be obtained from the symmetrical version of the classical two-time correlations, and expressed as
\begin{equation}
K_{j_{ij}j_{lm}}(\tau,0)=\frac{1}{2}\left\langle\left\{\hat{j}_{ij}(\tau)\,,\,\hat{j}_{lm}(0)\right\}\right\rangle-\left\langle\hat{j}_{ij}(\tau)\right\rangle\left\langle\hat{j}_{lm}(0)\right\rangle\,.
\label{Cj1}
\end{equation}
Theoretically, the response of a system under external perturbations can be studied in terms of these correlations functions \cite{weiss19991,rammer20071}. Notice that we evaluate the fluctuations in the non-equilibrium stationary state, so that we might expect that Eq.(\ref{Cj1}) can elucidate some properties of stationary non-equilibrium, rather than the equilibrium quantum correlations \cite{souza20081}. Furthermore, it has been shown that Eq.(\ref{Cj1}) is related to the fluctuations of the stationary energy current across the chain \cite{kundu20101}.

As the stationary state obeys a Gaussian distribution, the four-time correlation terms implicit in $K_{j_{ij}j_{lm}}(t,t')$ can be decomposed into terms involving the product of two-time correlations, in the form
\begin{eqnarray}
 &&\left\langle \left\{\left\lbrace \hat x_{i}(\tau),\hat p_{j}(\tau) \right\rbrace , \left\lbrace \hat x_{l}(0),\hat p_{m} (0)\right\rbrace\right\}\right\rangle = 2\,\left\langle \left\lbrace \hat x_{i}(\tau),\hat p_{j}(\tau) \right\rbrace\right\rangle   \left\langle \left\lbrace \hat x_{l}(0),\hat p_{m} (0)\right\rbrace\right\rangle  \nonumber \\
 &+&2\,\left[\left\langle \left\lbrace \hat x_{i}(\tau),\hat x_{l}(0) \right\rbrace\right\rangle  \left\langle   \left\lbrace \hat p_{j}(\tau),\hat p_{m} (0)\right\rbrace\right\rangle + \left\langle \left[ \hat x_{i}(\tau),\hat x_{l}(0) \right]\right\rangle  \left\langle   \left[ \hat p_{j}(\tau),\hat p_{m} (0)\right]\right\rangle\,\right] \nonumber \\
&+&2\,\left[\left\langle \left\lbrace \hat x_{i}(\tau),\hat p_{m}(0) \right\rbrace\right\rangle  \left\langle   \left\lbrace \hat x_{l} (0),\hat p_{j}(\tau)\right\rbrace\right\rangle - 
\left\langle \left[ \hat x_{i}(\tau),\hat p_{m}(0) \right]\right\rangle  \left\langle   \left[ \hat x_{l} (0) ,\hat p_{j}(\tau)\right]\right\rangle\right]
 \,.
\label{FTC}
\end{eqnarray} 
Then the current-current response function (\ref{Cj1}) can be expressed as
\begin{eqnarray}
K_{j_{ij}j_{lm}}(\tau,0)=\frac{k^2}{4m^2}\sum_{\substack{ \alpha,\alpha'=i,j \\ \beta,\beta'=l,m }}\left[\,S_{\alpha,\beta}\,\left[C_{x_\alpha x_\beta}(\tau,0)\,C_{p_{\alpha'}p_{\beta'}}(\tau,0)+Y_{x_\alpha x_\beta}(\tau,0)\,Y_{p_{\alpha'}p_{\beta'}}(\tau,0)\right]\right. \\ \nonumber
\left.\,+\,S_{\alpha,\beta'}\left[\,C_{x_\alpha p_\beta}(\tau,0)\,C_{x_{\beta'}p_{\alpha'}}(-\tau,0)-
Y_{x_\alpha p_\beta}(\tau,0)\,Y_{x_{\beta'}p_{\alpha'}}(-\tau,0)\right]
\,\right]\,,\hspace*{1cm}
\end{eqnarray}
where $S_{a,b}$ is the sign of the cofactor of the element $(a,b)$ in the $4\times 4$ array defined by the indexes $\{i,j,l,m\}$. Finally, according to Eq.(\ref{GEF1}), the autocorrelation function of the total current flowing from the ${\cal L}$- to the ${\cal R}$-oscillator is given by
\begin{equation}
K_{JJ}(\tau,0)=K_{j_{{}_{{\cal C}{\cal L}}} j_{{}_{{\cal C}{\cal L}}}}(\tau,0)+K_{j_{{}_{{\cal C}{\cal L}}} j_{{}_{{\cal R}{\cal C}}}}(\tau,0)+K_{j_{{}_{{\cal R}{\cal C}}} j_{{}_{{\cal C}{\cal L}}}}(\tau,0)+K_{j_{{}_{{\cal R}{\cal C}}} j_{{}_{{\cal R}{\cal C}}}}(\tau,0)\,.
\label{CCR}
\end{equation}

In contrast to the average energy current, the correlation function $K_{JJ}(\tau,0)$ involves crossed terms between the ${\cal R}$- and the ${\cal L} $- oscillators. In addition, while $\left\langle\hat{J}\right\rangle$ (\ref{TEC}) is given by a linear combination of two-time correlation terms, $K_{JJ}(\tau,0)$ has a nonlinear dependence on such terms. These two aspects will be useful in the subsequent discussion. Alternatively, the behavior of $\left\langle\hat{J}\right\rangle$ and $K_{JJ}(\tau,0)$ will help us to gauge whether the average properties of the energy current are sensitive to quantum correlations, such as genuine tripartite entanglement.


\subsection{Quantum Correlations: Discord and Entanglement}


We shall analyze the two-mode quantum correlations between the $ith$ and $jth$ oscillators by means of the discord measure on the right \cite{giorda20101,adesso20101}, denoted by $ D^{\leftarrow}(\hat \rho_{ij})$. The entanglement between both modes can be quantified by the well-known logarithmic negativity $E_{N}(\hat{\rho}_{ij})$ \cite{vidal20021,plenio20051,adesso20071}. In particular we devote special attention to the entanglement $E_{N}(\hat{\rho}_{{\cal L}{\cal R}})$ and the discord $ D^{\leftarrow}(\hat{\rho}_{{\cal L}{\cal R}})$ between the side oscillators. 

We use a recent criterion in the realm of continuous-variable systems \cite{valido20141} to study tripartite entanglement, which is a good estimator for $\kappa$-partite entanglement \cite{levi20131} in $n$-mode Gaussian as well as non-Gaussian states. A tripartite harmonic system may develop  \textit{bipartite} three-mode entanglement, which means that there is at least a bipartition of the three-mode system that is entangled, or \textit{genuine tripartite} entanglement, which corresponds to the case in which all the bipartitions are entangled and the state $\hat{\rho}_{{\cal L}{\cal C}{\cal R}}$ cannot be written as a convex combination of bipartite separable states. Here, the criterion reduces to evaluate a figure of merit ${\cal T}_{\kappa,n}$, such as a positive value of ${\cal T}_{3,3}(\hat{\rho}_{{\cal L}{\cal C}{\cal R}})$ (${\cal T}_{2,3}(\hat{\rho}_{{\cal L}{\cal C}{\cal R}})$) indicates that the state $\hat{\rho}_{{\cal L}{\cal C}{\cal R}}$ is genuine tripartite entangled (bipartite three-mode entangled) \cite{levi20131,valido20141}.

As we are dealing with stationary Gaussian states, all the previously mentioned indicators of quantum correlations can be directly computed from the covariance matrix $\vect V$ given by Eq.(\ref{CVmatrix}). The logarithmic negativity can be expressed as \cite{adesso20071}, 
\begin{equation}
\label{LogNeg} 
E_{N}\left(\hat\rho_{ij} \right)
= \mathop{\text{max}}\lbrace0,-\ln\left( 2\nu_{-}\right) \rbrace,
\end{equation}
where $\nu_{-}$ stands for the lowest symplectic eigenvalue of the partial transpose covariance matrix $\vect V_{ij}^{T_{j}}$, corresponding to the reduced density matrix $\hat\rho_{ij}$. The (right) discord is given by \cite{olliver20011,henderson20011},
\begin{equation}
D^{\leftarrow}\left(\hat\rho_{ij} \right)= I(\hat\rho_{ij})-{\cal I}^{\leftarrow}(\hat\rho_{ij}),
\label{disc}
\end{equation}
with the total correlations
\begin{equation}
I(\hat\rho_{ij})=S(\hat\rho_{i})+S(\hat\rho_{j})-S(\hat\rho_{ij}) 
\end{equation}
and the classical correlations
\begin{equation}
{\cal I}^{\leftarrow}(\hat\rho_{ij})=\displaystyle\max_{\Pi^{(j)}_{l}}\left\lbrace S(\hat\rho_{ij})-\sum_{l} p_{l} S(\hat\rho_{i}^{(l)})\right\rbrace\,,
\label{disc2}
\end{equation} 
which are given in terms of the von Newman entropy $S(\hat\rho)$. Closed form expressions for the quantum discord as a function of the covariance matrix $\vect V_{ij}$ have been derived in \cite{giorda20101,adesso20101}. It is important to realize that these indicators of quantum correlations involve a non-linear dependence on the density operator and the two-time correlation functions.  See Appendix \ref{ASC} for further details of the logarithmic negativity, the quantum discord and the separability criteria ${\cal T}_{\kappa,n}$.


\section{Results}\label{SRES}


We now investigate the average properties of the total current $\hat{J}$ when the three oscillators share two-mode and tripartite entanglement. In many realistic situations, e.g. quantum Brownian motion, the interaction with the environment leads to an Ohmic dissipation. In a first approach, for a nanomechanical setup, one may think that the thermal relaxation is mainly due to the coupling with the acoustic phonons of the substrate, which may lead to linear spectral density at low oscillator frequencies. However, in some cases, the dimensionality of the environment may induce super-Ohmic dissipation. Here we analyze both the Ohmic and super-Ohmic dissipations, which are characterized by the spectral densities
\begin{equation}
J^{(Oh)}_{i}(\omega)=\frac{\pi m\gamma_{i}}{2}\,\omega \, e^{-\omega/\omega_{c}}, \label{SD1} 
\end{equation}
and
\begin{equation}
J^{(SOh)}_{i}(\omega)=\frac{\pi m\gamma_{i}}{2}\,\frac{\omega^2}{\omega_{c}} \, e^{-\omega/\omega_{c}}, \label{SD2}
\end{equation}
respectively, with $\gamma_{i}$ the dissipative rate for the $ith$ oscillator and $\omega_{c}$ the frequency cut-off of the environmental spectrum. As argued in Section \ref{SLA}, the stationary state is reached in a time scale larger than any of the natural time scales implicit in the dynamics of the open chain; namely $\left\lbrace \omega_{c}^{-1},\gamma^{-1},\hbar/2\pi K_{B} T\right\rbrace $. 

From now on we set the environmental parameters $\gamma_{1}=\gamma_{3}=10^{-4}\Omega$, $\gamma_{2}=0.05\Omega$, $\omega_{c}=20 \Omega$, and the typical values for nanomechanical oscillators $\Omega=1\,GHz$ and $m=10^{-16}\,kg$. With this configuration the system begins to exhibit quantumness at temperatures in the range of mili-Kelvin. We also assume off resonance oscillators with frequencies, $\omega_{{\cal L}}=\Omega+0.4\,\delta \omega$, $\omega_{{\cal C}}=\Omega+0.9\,\delta\omega$, and $\omega_{{\cal R}}=\Omega-0.7\,\delta\omega$, given in terms of a detuning parameter $\delta \omega$. 


\subsection{Two-mode entanglement and average energy current}\label{STEEC}


\begin{figure*}[ht!]
\begin{center}
\includegraphics[scale=0.16]{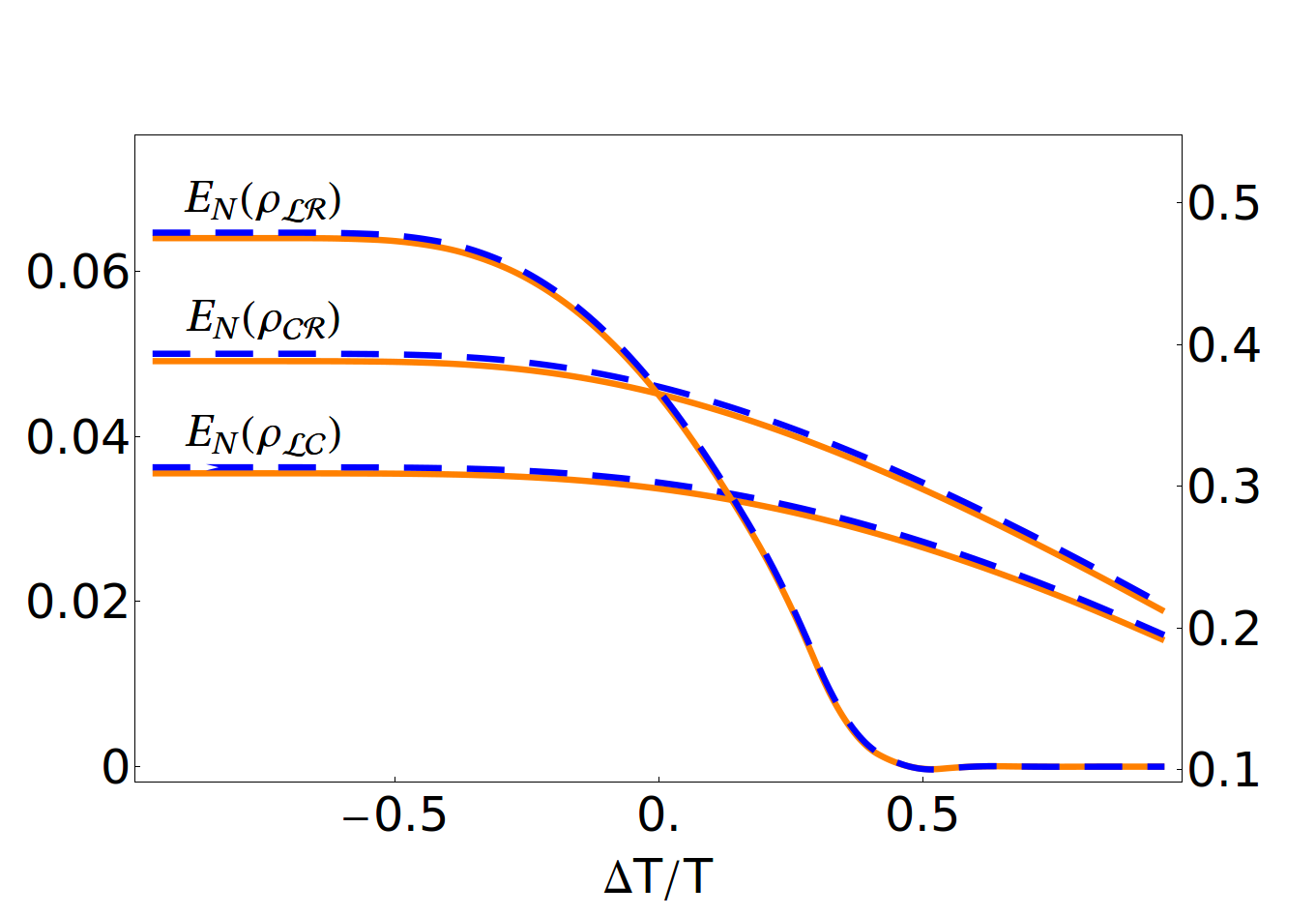}
\includegraphics[scale=0.19]{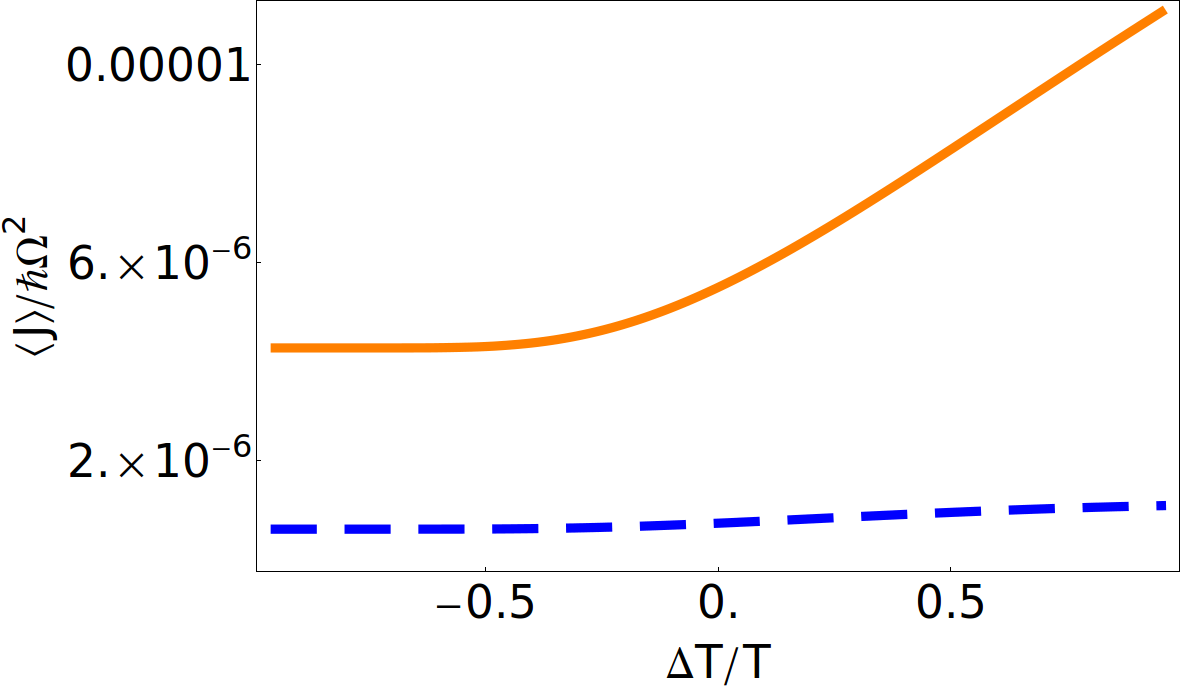}
\caption{(color online). 
Left: The two-mode ${\cal L}|{\cal R}\,-$ entanglement (labels on the left), and the ${\cal C}|{\cal R}\,-\,,\,\,{\cal L}|{\cal C}\,-$ entanglements (labels on the right) as a function of the temperature gradient $\Delta T$. Right: The average of the total energy current across the chain as function of the temperature gradient. On both panels the orange solid line corresponds to Ohmic dissipation, and the blue dashed line to super-Ohmic dissipation. The temperature gradient must satisfy $\Delta T/T>-1$ to prevent the temperature of the central oscillator becomes negative. We have fixed $\delta\omega/\Omega=0.5$, $k/m \Omega^2=1.8$, $\delta T/T=0.95 $ and $k_{B}T/\hbar \Omega\simeq 0.27$. \label{Fig1}} 
\end{center}
\end{figure*}

We start by analyzing the behavior of two-mode entanglements and the total energy current with the temperature gradient $\Delta T$. As figure \ref{Fig1}) shows, the three oscillators become two-mode entangled at low temperature gradients, with this entanglement exhibiting a plateau for negative gradients. Interestingly the total current flowing through the oscillator chain presents a similar plateau. This can be related to the proximity of the central oscillator to its ground state at very low temperatures $T_{\cal C}$, which is effectively reached for $\Delta T/T\simeq -1$. A similar result for entanglement has been obtained in the study of the influence of heat transport on the two-mode entanglement between oscillators that are embedded in a disordered harmonic chain connected to heat baths at both ends \cite{gaul20071}.  
It has been shown that a plateau emerges when the energy spectrum is bounded from below since each site of the chain suffers an harmonic potential. As the central oscillator gets closer to the ground state for negative values of $\Delta T$, the energy flowing across this oscillator becomes bounded as the temperature gradient decreases. In the absence of the harmonic confinement the logarithmic negativity would continue growing up to a maximum value, as the heat transport decreases \cite{gaul20071}. Moreover, the plateau in the entanglement remains even when the average energy current across the chain becomes zero, though the temperature gradient $\Delta T$ is not zero. This occurs when $\delta T=0$ and the left and right oscillators are identical $\Omega_{{\cal L}}=\Omega_{{\cal R}}$, and therefore $\left\langle \hat{j}_{{\cal C}{\cal L}}\right\rangle=-\left\langle \hat{j}_{{\cal R}{\cal C}}\right\rangle$. This last point also underlines that the appearance of entanglement is mainly attributed to proximity of the system to its ground state, rather than to the presence of an energy flow induced by non-equilibrium conditions.

The two-mode entanglement rapidly decreases for positive $\Delta T$, while the energy current grows monotonically. This is expected as the temperature of the whole three-mode system is increased on average, which is generally harmful for entanglement. Previous results have suggested this behavior, in fact it has been shown that in a harmonic chain an increasing $\delta T$ is detrimental to build up bipartite or tripartite entanglement due to the rise in the thermal noise \cite{valido20131}. In addition, it can be shown from (\ref{CFFDAL}) that an initially squeezed environmental state effectively increases the temperature. Hence, in the present setting an initial squeezed bath state does not favor the appearance of stationary entanglement. 

Moreover, one may expect that non-Markovian effects, which are more relevant for super-Ohmic dissipation, would substantially degrade the entanglement with respect to the Ohmic situation. According to figure (\ref{Fig1}), the two-mode entanglement is essentially the same for the chain suffering Ohmic or super-Ohmic dissipation; namely, $E_{N}(\hat{\rho}_{{\cal L}{\cal R}})$, $E_{N}(\hat{\rho}_{{\cal R}{\cal C}})$ and $E_{N}(\hat{\rho}_{{\cal C}{\cal L}})$ practically coincide for both situations. This result is in contrast with the observed transient evolution of the two-mode entanglements under different environmental spectral densities, in which the super-Ohmic dissipation induces stronger disentanglement effects \cite{an20071}. The coincidence of the stationary two-mode entanglements also differs from the emergence of entanglement in a situation in which the oscillators are affected by the same bath \cite{valido20132}. In the case of the energy current, we observe that it is strongly affected by the interaction with the heat baths, determined by the fixed spectral density.
  
\begin{figure*}[ht!]
\begin{center}
\includegraphics[scale=0.25]{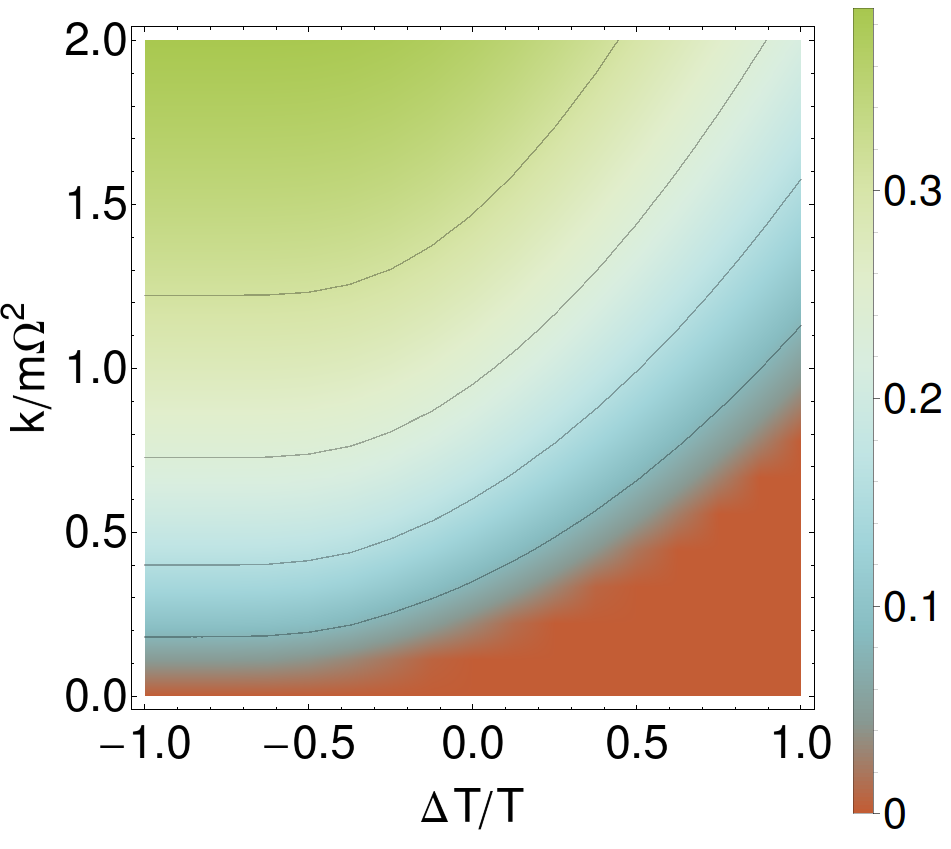}
\includegraphics[scale=0.25]{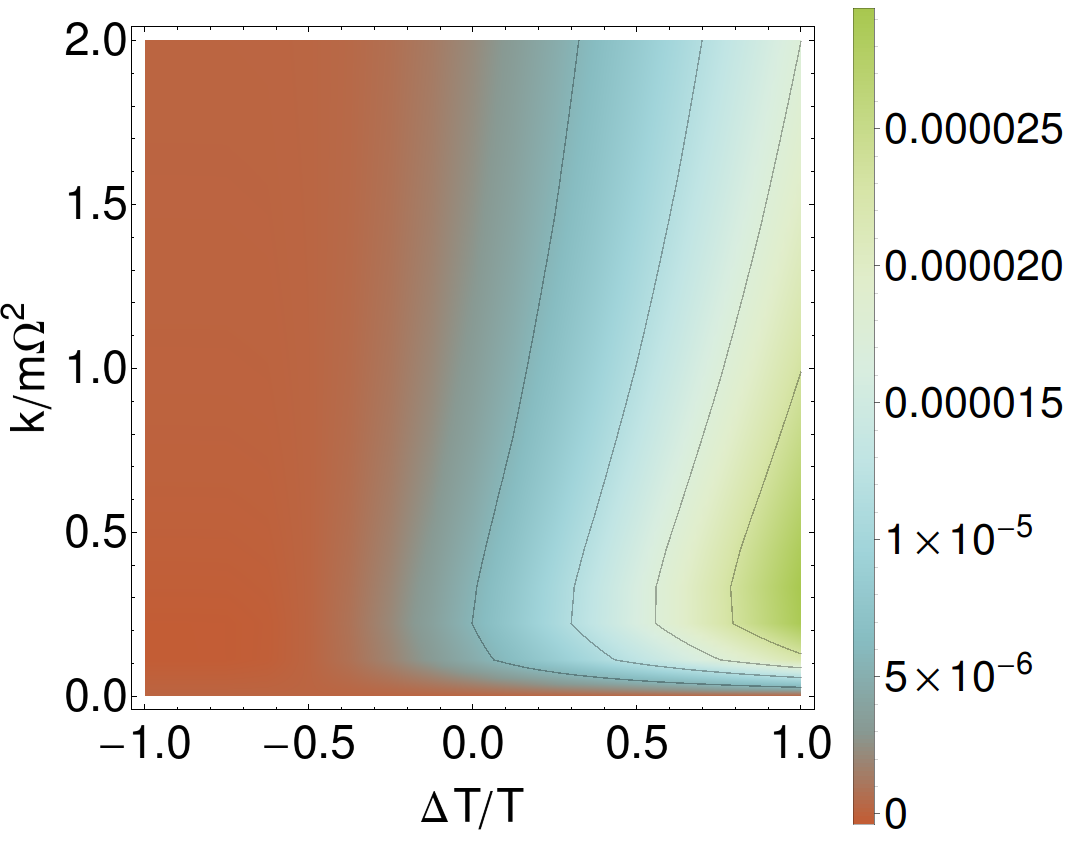}
\caption{(color online). The two-mode entanglement $E_{N}(\hat{\rho}_{{\cal R}{\cal C}})$ (left panel) and the stationary energy current $\left\langle\hat{j}_{{\cal R}{\cal C}}\right\rangle/\hbar \Omega^{2}$ (right panel) in terms of the temperature gradient $\Delta T$ and the coupling strength $k$ under Ohmic dissipation. The system parameters are the same as figure (\ref{Fig1}).\label{Fig2}}
\end{center}
\end{figure*}

We have performed an extensive analysis of the two-mode entanglements and the energy currents involving the central oscillator, in terms of both the temperature gradient $\Delta T$ and the coupling strength $k$. As figure (\ref{Fig2}) shows, a similar behavior to that illustrated in the figure (\ref{Fig1}) is reproduced for different values of $k$. The entanglements $E_{N}(\hat{\rho}_{{\cal R}{\cal C}})$ and $E_{N}(\hat{\rho}_{{\cal L}{\cal C}})$ increase for lower temperature gradients and stronger couplings. Whereas the energy currents $\left\langle \hat{j}_{{\cal R}{\cal C}}\right\rangle $ and $\left\langle \hat{j}_{{\cal C}{\cal L}}\right\rangle$ exhibit a relatively weak dependence on the coupling strength, and the expected increase with the temperature gradient. Also the plateau of small energy currents arising in the proximity of the ground state of the central oscillator can be clearly observed.

Hence, our results indicate that the energy currents across the system are insensitive to the emerge of two-mode entanglements between the oscillators, both under Ohmic and super-Ohmic dissipation, and irrespective of the coupling strength with the heat baths. The two-mode entanglement $E_{N}(\hat{\rho}_{{\cal L}{\cal R}})$ and total energy current $\left\langle \hat{J}\right\rangle$ remain nearly unchanged provided that the central oscillator is close enough to the ground state, at temperatures between $T_{\cal{L}}$ and $T/2$. An increase in the temperature of this oscillator results in a deterioration of the entanglement, and an increase in the energy current.


\subsection{Energy current correlations and three-mode entanglement} \label{SCCR}


\begin{figure}[h]
\centering
\begin{center}
\includegraphics[scale=0.263]{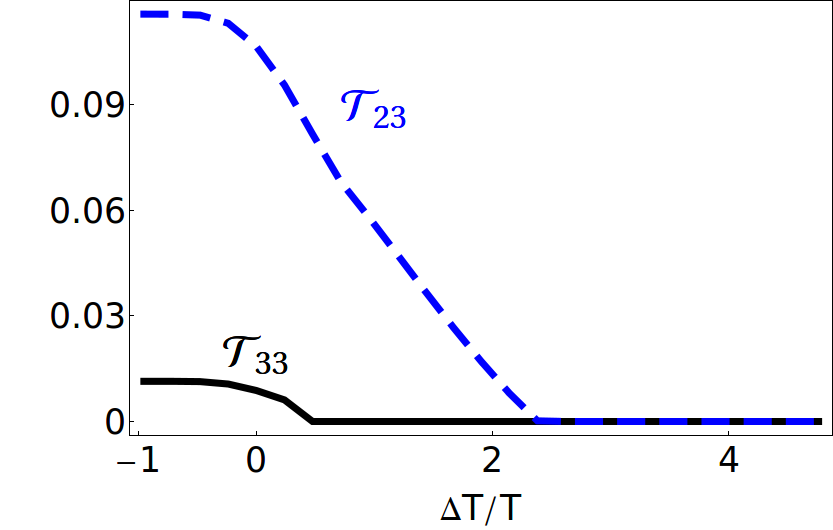} 
\includegraphics[scale=0.23]{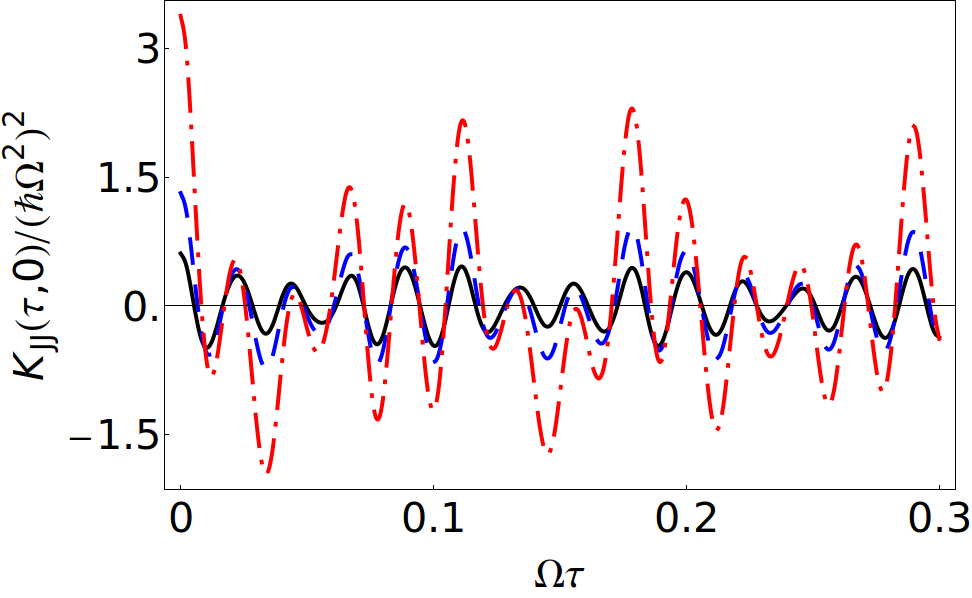} 
\end{center}
\caption{(color online). Left: The criteria ${\cal T}_{23}$ and ${\cal T}_{33}$ as a function of the temperature gradient for Ohmic dissipation. Right: The time evolution of the fluctuations of the total current under Ohmic dissipation, in a system that is genuine entangled $(\Delta T/T=-\,0.95)$ (black solid line), bipartite three-mode entangled $(\Delta T/T=\,1.9)$ (blue dashed line), and likely separable in the three possible bipartitions $(\Delta T/T=\,4.3)$ (red dot-dashed line). The remaining parameters are $k/m \Omega^2=2$, $\delta\omega/\Omega=0.5$, $\delta T/T=0.95$, and $k_{B}T/\hbar \Omega\simeq 0.27$.}
\label{Fig3}
\end{figure}

In this section we analyze the three-mode entanglement and the energy current correlations between the left and right oscillators, which includes correlation terms involving the three oscillators, see Eq.(\ref{CCR}). Figure (\ref{Fig3}) shows the bipartite three-mode ($\kappa=2$) and the genuine tripartite ($\kappa=3$) entanglements measured by the corresponding criteria ${\cal T}_{\kappa,3}$ \cite{valido20141,levi20131}. The results for both the Ohmic and super-Ohmic dissipations are quite similar. In the low temperature and strong coupling regime, the three-mode system exhibits genuine tripartite entanglement, though this feature rapidly disappears for positive values of $\Delta T$, such as occurs with the two-mode entanglement. Strikingly, the system still remains bipartite three-mode entangled at relatively high temperature gradients ($\Delta T/T\approx 2$). Hence the tripartite entanglement is more robust to temperature changes than the two-mode entanglement between the side oscillators.

Figure (\ref{Fig3}) also shows the initial time evolution of the energy current correlations for three different three-mode entanglement configurations. As expected, the fluctuations of the energy current exhibit an oscillatory behavior, which should be progressively attenuated at larger time intervals. According to a previously reported exponential time decay of the two-time correlation functions (\ref{SLEq1}) in a damped harmonic oscillator at low temperature $T$, such oscillations should be effectively suppressed at time $\tau>\hbar/2\pi k_{B}T$ \cite{jung19851}.

As evidenced by figure (\ref{Fig3}), the energy current correlations exhibit similar oscillations as the system evolves from genuine tripartite to bipartite three-mode entanglement. The most significant discrepancy between these two configurations is an increase in the oscillation amplitude, which can be attributed to the thermal fluctuations that arise with increasing the temperature gradient. Indeed, a similar oscillating behavior in the fluctuations is still observed at relatively large temperature gradients $(\Delta T/T\gtrsim 4)$, when the system is expected to be separable in the three possible bipartitions. 

The results we have obtained from an analysis considering an extensive set of parameters $\left\lbrace T,\,k,\,\delta \omega,\,\delta T\right\rbrace$ corresponding to different multipartite entanglement configurations, for both Ohmic and super-Ohmic dissipations, also indicate that the energy currents correlations across the harmonic chain are insensitive to the emerge of tripartite genuine or bipartite three-mode entanglement.

\begin{figure*}[ht!]
\begin{center}
\includegraphics[width=0.35\textwidth]{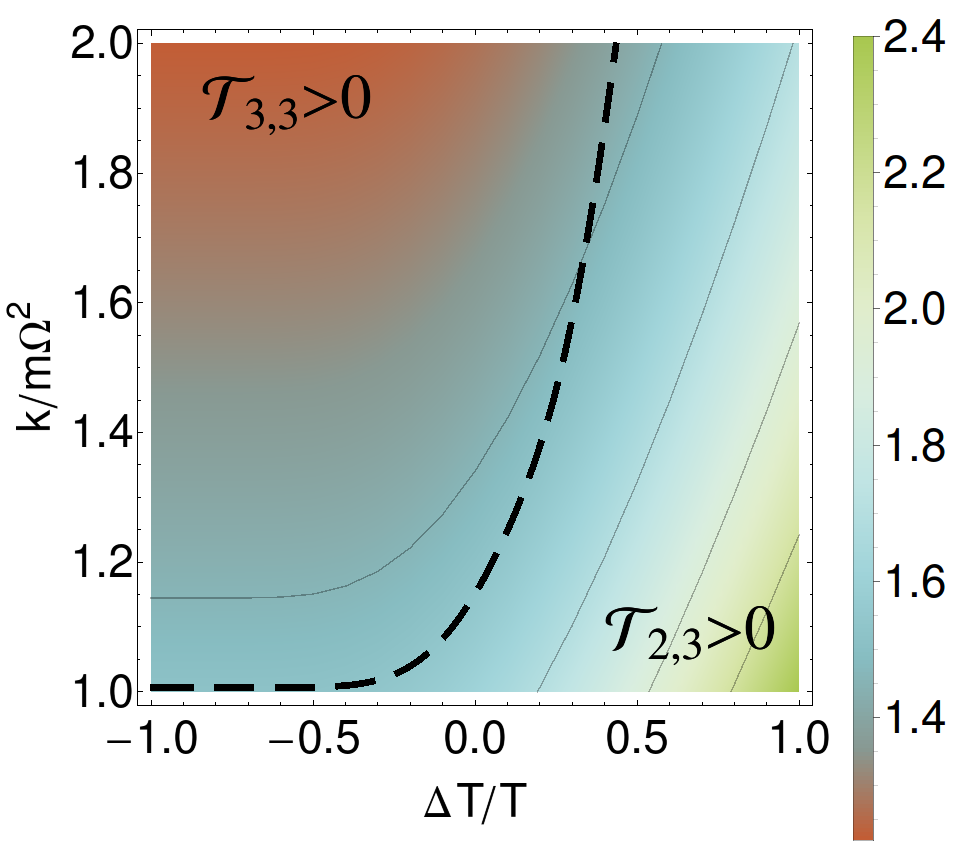}
\caption{(color online). Density plot of the energy current correlations $K_{JJ}(t,t)/(\hbar\Omega^2)^2$ as a function of the temperature gradient $\Delta T$ and the coupling strength $k$ for Ohmic dissipation. A similar result is obtained for super-Ohmic dissipation. The black dashed line delimits the states in which the hierarchy ${\cal T}_{3,3}$ changes from positive to negative. The parameters are the same as in figure (\ref{Fig3}).\label{Fig31}} 
\end{center}
\end{figure*}

To conclude this section we focus on the energy current correlations evaluated at equal time, see figure (\ref{Fig31}), which will be useful in the subsequent discussion. Both the stationary fluctuations and the average of the energy current, see figure (\ref{Fig2}), grow with increasing the temperature gradient. But, in contrast to the average energy current, the plateau at small values of the fluctuations is observed above a given value of the coupling strength, which is larger in the case of Ohmic dissipation.

Once again, the fluctuations of the energy current are insensitive to whether the system experiences bipartite three-mode or genuine tripartite entanglement. Similar results are obtained for the current-current response involving the central and the side oscillators.


\subsection{Quantum Discord}\label{SNC}


One might expect that a scenario similar to the one previously described for the two- and three-mode entanglement would be repeated in the presence of other non-classical correlations, such as discord. In this section we analyze a possible connection between the energy current and the quantum correlations measured by the right-discord $D^{\leftarrow}(\hat{\rho}_{{\cal R}{\cal L}})$. Although not shown in this work, similar results are obtained from the analysis of the discord $D^{\rightarrow}(\hat{\rho}_{{\cal R}{\cal L}})$ measured from the left . We also point out that the two-mode discord contains the contribution of the two-mode entanglement studied in previous sections.

\begin{figure*}[ht!]
\begin{center}
\includegraphics[scale=0.11]{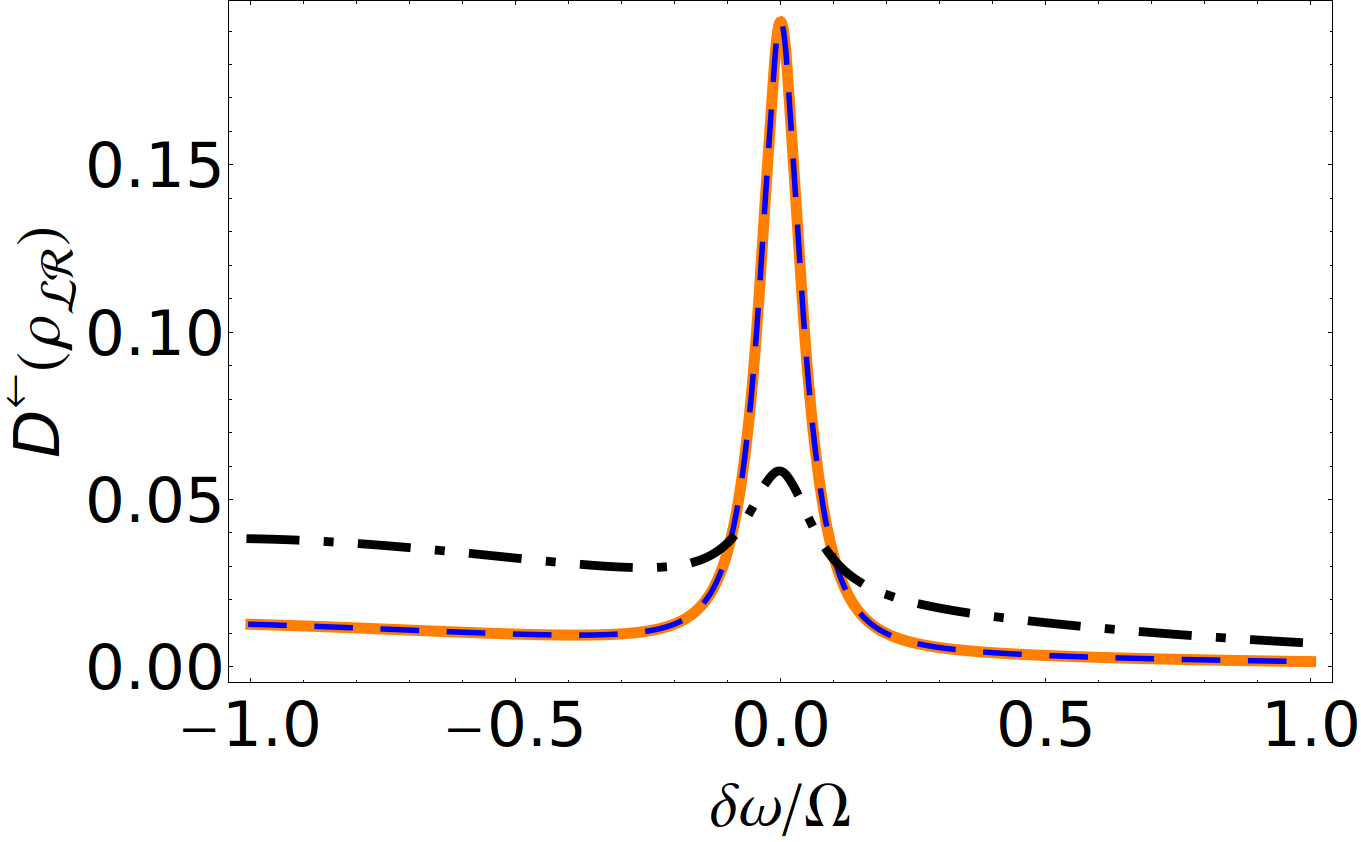}
\includegraphics[scale=0.11]{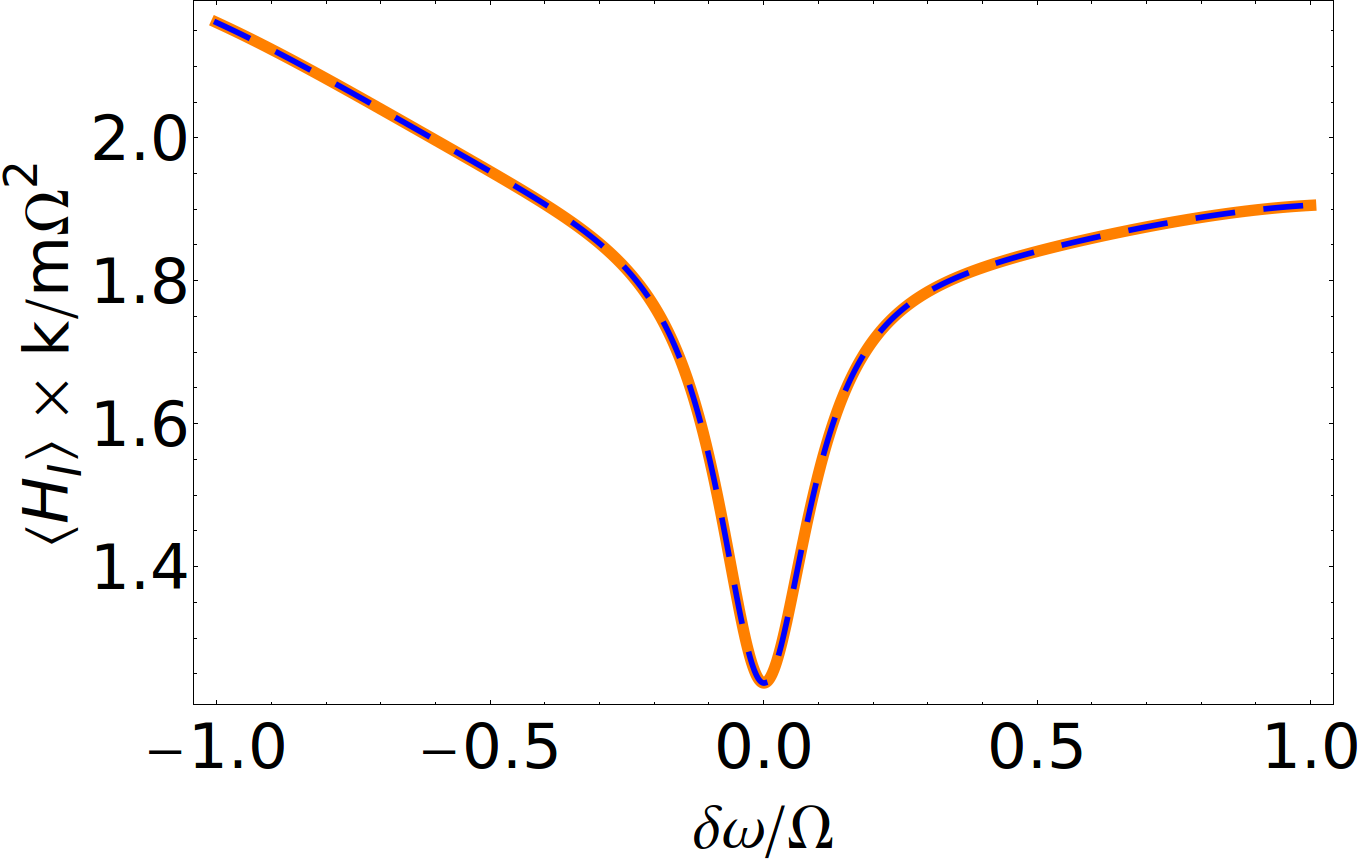}
\includegraphics[scale=0.11]{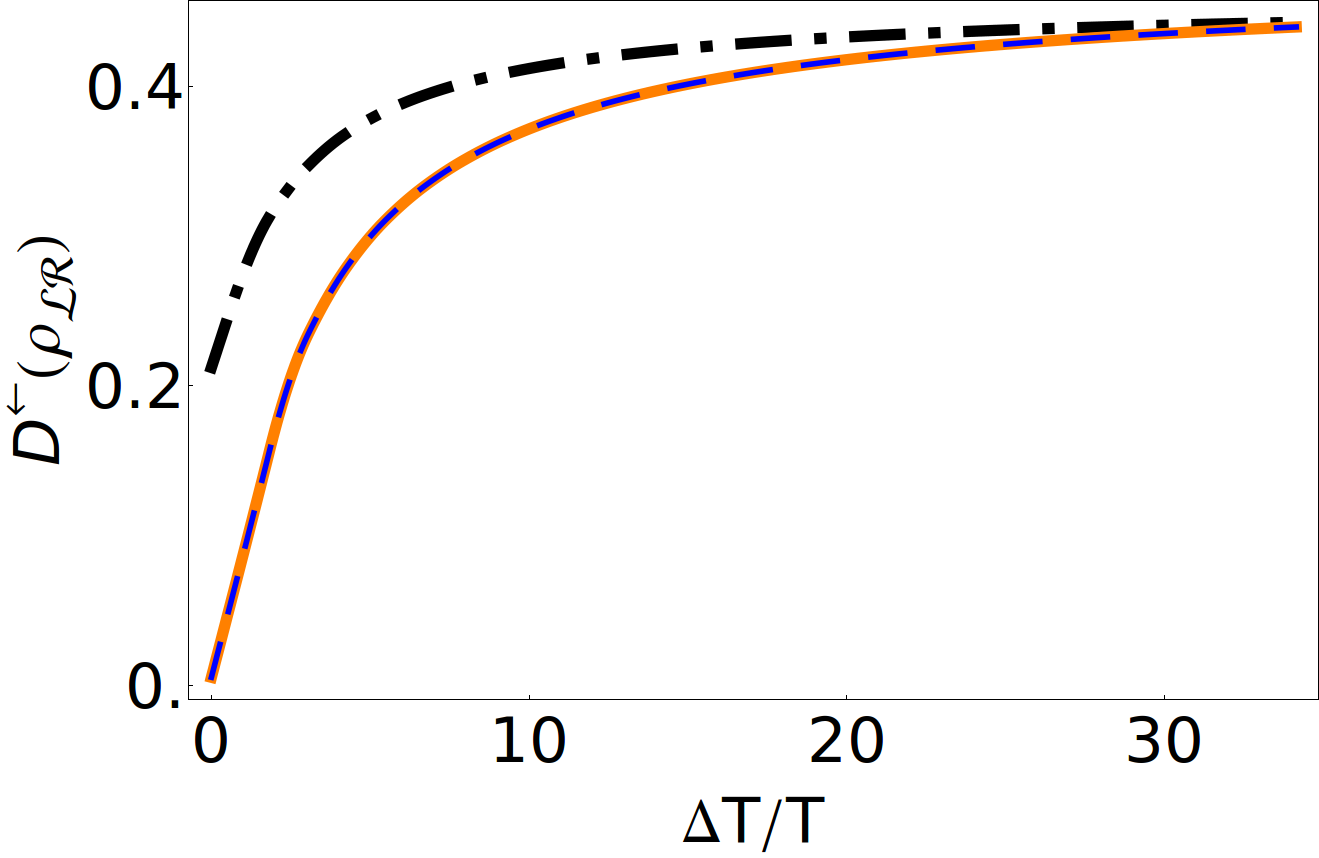}
\caption{(color online). Left: The right-discord as a function of the detuning $\delta\omega$, for the temperature gradients $\Delta T/T=3.8$ and $0.95$ (black dot-dashed line). Center: The averaged interaction energy $\hat H_{I}$, see Eqs. (\ref{hamiltonian}) and (\ref{MEI}), in terms of $\delta\omega$. Right: The right-discord at resonance $(\delta\omega=0)$, as a function of the temperature gradient. The black dot-dashed line gives both the Ohmic and super-Ohmic dissipative discord for an initially squeezed central reservoir, with $r_{{\cal C}}=1\,(r_{{\cal R}}=r_{{\cal L}}=0)$. In the three panels the orange-solid and blue-dashed lines correspond to Ohmic and super-Ohmic dissipations respectively, and the parameters are $k/m \Omega^2=1.8$, $\delta T/T=0.95$, and $k_{B}T/\hbar \Omega\simeq 0.53$.\label{Fig4}} 
\end{center}
\end{figure*} 

We have found that discord exhibits a strong dependence on both the temperature gradient and the frequencies of the oscillators. As shown in figure (\ref{Fig4}), its presents a sharp peak centered at the resonance condition $(\delta \omega = 0)$, and with an amplitude that grows with increasing the temperature gradient. Interestingly, the discord appears strongly correlated with the mean interaction energy between the oscillators,
\begin{equation}
\left\langle \hat H_{I}\right\rangle =\frac{k}{2}\Big[C_{x_{{\cal L}}x_{{\cal L}}}(t,t)+2\,C_{x_{{\cal C}}x_{{\cal C}}}(t,t)+C_{ x_{{\cal R}} x_{{\cal R}}}(t,t)-2\underbrace{\left(\,C_{x_{{\cal L}}x_{{\cal C}}}(t,t)+C_{x_{{\cal R}}x_{{\cal C}}}(t,t)\,\right)}_{\text{\normalsize{Correlation terms}}}\Big]\,,
\label{MEI}
\end{equation}
which has maximum strength also at resonance, see figure (\ref{Fig4}). This resonant interaction becomes stronger for higher temperature gradients, which also increase the discord. As all the negative contribution to $\left\langle \hat H_{I}\right\rangle$ comes from the crossed correlation terms, it becomes evident that the maximum interaction strength occurs when these correlations take the highest values, which also turns into the optimal conditions for discord. This result evidences an underlying connection between the interaction energy and the discord, as could be anticipated considering that the discord would be zero in the absence of interaction.

Figure (\ref{Fig4}) also shows that the discord in the resonant system begins to grow almost linearly with the temperature gradient, and then approaches a constant value at higher gradients. An initially squeezed central reservoir enhances the creation of discord, both for Ohmic and super-Ohmic dissipation. This is in agreement with the foregoing results, as the squeeze of the initial bath state effectively increases the temperature perceived by the oscillators, see Eq.(\ref{CFFDAL}). Then an increase of the  stationary discord between the side oscillators may be induced either by initially squeezing the central reservoir or increasing its temperature. It has been shown that discord may be additionally created by local noisy operations, such as dissipation \cite{wu20111,streltsov20111}. Hence, it may happen that discord would be generated by an energy current induced by a temperature gradient, as this current would make each oscillator to dissipate.

Considering that the discord contains all the quantum correlations, it would be interesting to analyze whether the entanglement available in the system contributes to its increase with the temperature gradient. At this respect, since entanglement can be only created by non-local manipulations and it becomes zero at high temperature gradients, see Section (\ref{STEEC}), we may conclude that such increase of the discord must be mainly due to local operations. The paramount role of the local manipulations in the creation of quantum correlations at resonance conditions is correlated with a maximum average interaction strength between oscillators of similar frequencies, see figure (\ref{Fig4}) and Eq. (\ref{MEI}). Notice that the discord grows with $\Delta T$ even in the absence of an energy current between the side oscillators ($\delta T=0$). 

The plateau of maximum discord at high temperature gradients can be attributed to the very low temperatures of the side oscillators $T_{{\cal L},{\cal R}}<\hbar \Omega/k_B$, which guarantees the ``coherence'' of the local manipulations. Indeed, the increase in the discord gradually disappear as the mean temperature $T$ increases, and therefore the three-mode system approaches to a classic state. We have observed that the discord $D^{\leftarrow}(\hat{\rho}_{{\cal R}{\cal L}})$ has almost disappeared at temperature $T\simeq 50\,\hbar \Omega/k_B$.

\begin{figure}[ht!]
\begin{center}
\includegraphics[scale=0.25]{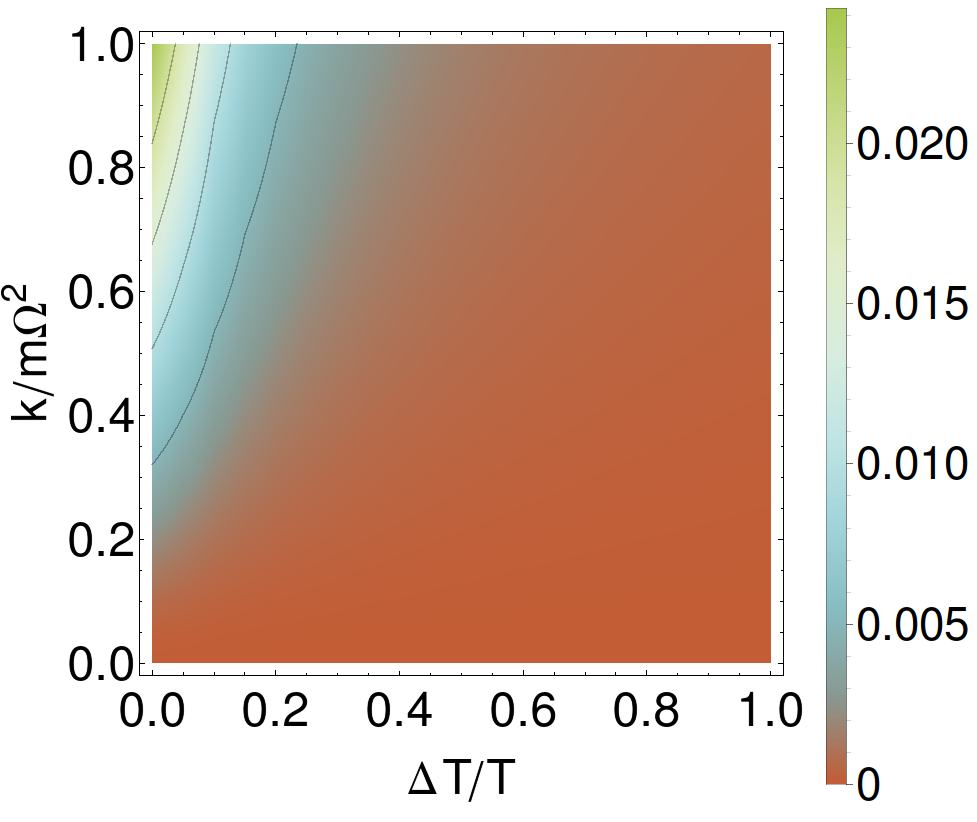}
\includegraphics[scale=0.25]{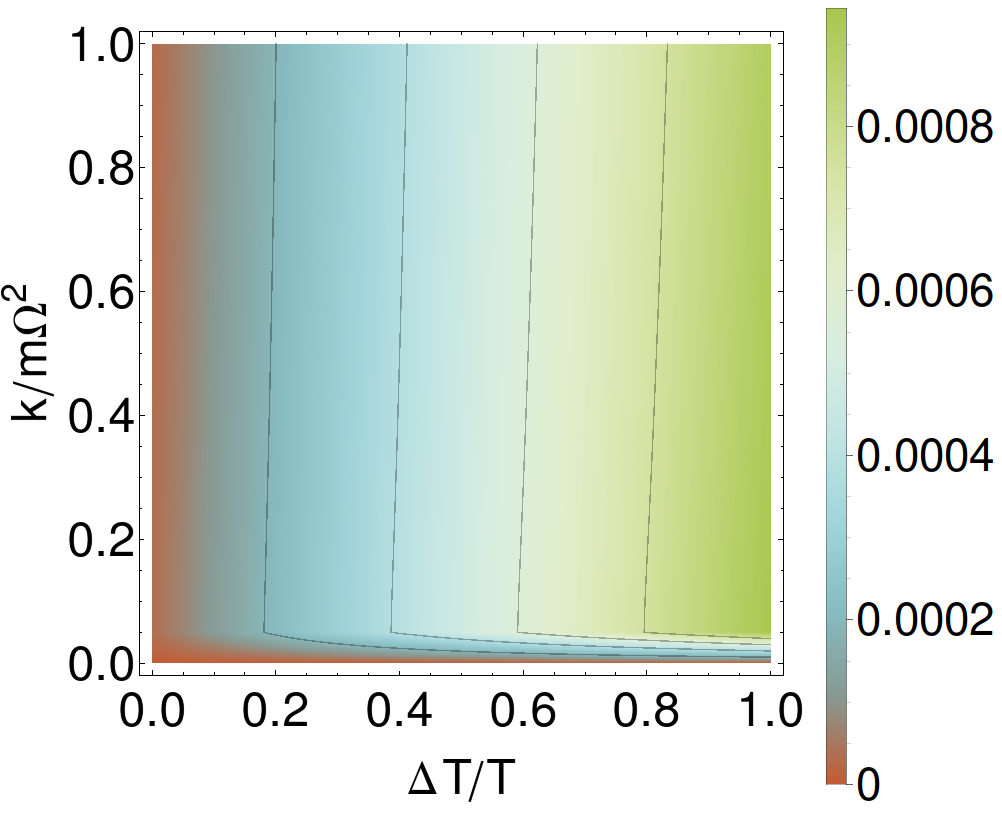}
\caption{(color online). The discord $D^{\leftarrow}(\hat{\rho}_{{\cal R}{\cal C}})$ (left panel) and the energy current $\left\langle \hat{j}_{{\cal R}{\cal C}}\right\rangle/\hbar \Omega^{2}$ (right panel) between the central and right oscillators in the resonant system under Ohmic dissipation. Similar results are obtained for the discord and the energy current between the ${\cal C}$- and ${\cal L}$- oscillators, both under Ohmic and super-Ohmic dissipations. The parameters are the same as figure (\ref{Fig4}).\label{Fig5}} 
\end{center}
\end{figure}

As expected, the two-mode discord between the central and the side oscillators is enhanced by increasing the interaction strength, see figure (\ref{Fig5}). Whereas, the trend of the central oscillator towards a classical state, by increasing its temperature through higher values of $\Delta T$, causes a progressive deterioration of the discord. In the case of the energy currents between the central and the side oscillators, they exhibit an almost linear increase with the temperature gradient, which is barely modified by the strength of the coupling interaction. Once again, the energy currents do not seem to be related to the significant non-classical correlations shared by the oscillators.

\section{Discussion}\label{SDIS}


Considering that the energy current between two oscillators has an explicit dependence on crossed correlations between them, one could expect that the emerge of entanglement in the system should have detectable effects on such current. Therefore, it would be interesting to determine whether a formal connection between entanglement and the average properties of the energy current can be formally established.

The results presented in Section \ref{STEEC} indicate that the behavior of the energy current is not modified by the presence of two-mode entanglement. A direct comparison between the two-mode entanglement between the side oscillators $E_{N}(\hat{\rho}_{{\cal R}{\cal L}})$
and the total energy current $\left\langle \hat{J}\right\rangle$ suggests an elusive correlation between them, see figure \ref{Fig1}). 

Although the most important contribution to the local currents $\left\langle \hat{j}_{ij}\right\rangle$ (\ref{EF1}) comes from crossed correlation terms that encode part of the quantum correlations shared by the oscillators, the total current $\left\langle \hat{J}\right\rangle$ (\ref{TEC}) does not have an explicit dependence on the correlation terms involving the ${\cal L}$- and ${\cal R}$- oscillators. Then the two-mode entanglement $E_{N}(\hat{\rho}_{{\cal R}{\cal L}})$ should not necessarily affect the total current. This reasoning does not exclude, however, the possibility that the total current could be sensitive to the two-mode entanglement shared by the central and the side oscillators.

The conjecture that entanglement and energy current are intimately related would lead to the natural question whether the current $\hat{j}_{ij}$ could serve as an useful witness of the entanglement between the $ith$ and $jth$ oscillators. According to the theory of entanglement, an entanglement witness $\hat{W}_{\hat{O}}$ based on a (bounded) Hermitian operator $\hat{O}$ may be constructed as
$\hat{W}_{\hat{O}}=\hat{O}-\text{inf}\left\lbrace\bra{\Psi_{i}}\bra{\Psi_{j}}\hat{O}\ket{\Psi_{i}}\ket{\Psi_{j}}\right\rbrace \mathbb{\hat I}$, where the last term is the infimum value of $\hat{O}$ among all the product states $\ket{\Psi_{i}}\ket{\Psi_{j}}$ \cite{sperling20091}. For $\hat{O}={\hat{j}_{ij}}$ (\ref{J1}) it follows
\begin{equation}
\hat{W}_{\hat{j}_{ij}}=\hat{j}_{ij}-\frac{k}{2m}\text{inf}\left\lbrace C^{Prod}_{x_{j}p_{j}}(t,t)-C^{Prod}_{x_{i}p_{i}}(t,t)+C^{Prod}_{x_{j}p_{i}}(t,t)-C^{Prod}_{x_{i}p_{j}}(t,t)\right\rbrace \mathbb{\hat I},
\label{EWJ}
\end{equation}
where $C^{Prod}_{ab}(t,t)$ is obtained from (\ref{SLEq1}) by considering the average over product states. $\hat{W}_{\hat{j}_{ij}}$ is a good candidate to unveil the two-mode entanglement $E_{N}(\hat{\rho}_{ij})$ provide that $\text{Tr}(\hat{\rho}_{ij}\hat{W}_{\hat{j}_{ij}})$ takes a negative value for at least one entangled state. Unfortunately such a rigorous proof requires a closed form expression for $C^{Prod}_{ab}(t,t)$, which is currently out of scope as we are dealing with a system under the non-equilibrium conditions induced by two different temperature gradients. Though we cannot guarantee whether Eq.(\ref{EWJ}) constitutes a good estimator of entanglement, the results of Section \ref{STEEC} evidence the difficulty of assessing entanglement through $\hat{W}_{\hat{j}_{ij}}$, mainly due to the apparent insensitivity of the energy current $\left\langle \hat{j}_{ij}\right\rangle $ to the two-mode entanglement $E_{N}(\hat{\rho}_{ij})$, see figure (\ref{Fig2}).

In addition, though the correlation terms in Eq. (\ref{J1}) partially carry the quantum correlations shared by the chain oscillators, they themselves do not necessarily manifest entanglement. Indeed, the so-called Peres-Hodorecki-Simon inequality \cite{simon20001}
\begin{equation}
\hat{\rho}_{ij} \ \text{entangled}\ \Longrightarrow  \ C_{x_{i}x_{j}}(t,t)C_{p_{i}p_{j}}(t,t)-C_{x_{i}p_{j}}(t,t)C_{p_{i}x_{j}}(t,t)< 0\,,
\label{SC}
\end{equation}
which provides a criterion to detect the two-mode entanglement between the $ith$ and $jth$ oscillators, already displays a non-linear relation between entanglement and the elements of the covariance matrix. 

A common feature of the entanglement measures (or entanglement monotones) is their non-linear functional dependence on the density operator \cite{guhne20091}, as occurs in the case of the logarithmic negativity \cite{vidal20021}. In some sense, the reliable observation of entanglement relies on the ability to measure non-linear properties of the quantum state \cite{mintert20071}. According to the expressions for the total energy current $\left\langle\hat{J}\right\rangle$ (\ref{EF1}) and the criterion for entanglement (\ref{SC}), the first-moment of the energy current depends linearly on the crossed correlation terms, whereas the entanglement exhibits a non-linear dependence on such terms. Hence, the energy current between the $ith$ and $jth$ is not expected to manifest the emergence of two-mode entanglement. 

Following the previous argument, the fluctuations of the total current (\ref{CCR}) could manifest the emergence of entanglement, as it involves correlation terms between all the oscillators. In Section \ref{SCCR} we focused on the tripartite entanglement, and showed that $K_{JJ}(t,t)$ seems to be insensitive to the inseparability properties of the three-mode chain. A similar conclusion was drawn from the comparison of the time evolution of the fluctuations $K_{JJ}(\tau,0)$ at different entanglement configurations of the stationary state, see figure (\ref{Fig3}). We remark that this result is not in contradiction with the previous argument based on criterion (\ref{SC}), as it provides a necessary, but not sufficient, condition for the existence of entanglement. 

Considering that, in contrast to entanglement, almost any quantum state has a non-negative discord \cite{ferraro20101}, a distinct behavior of these two quantum correlations might be expected \cite{horodecki20091}. The results of Section (\ref{SNC}) indicate that in the proximity of a resonance condition, a finite energy current induced by the temperature gradient $\Delta T$ may generate non-classical correlations between the left and right oscillators, even when the total energy current between them becomes zero. This behavior has been correlated with a maximum strength of the average interaction between the harmonic oscillators, see figure (\ref{Fig4}). The same results also show that the energy current is not modified by the emerge of discord, see figure (\ref{Fig5}). The average properties (mean value and fluctuations) of the energy current as a whole exhibit a `linear' behavior ruled by the temperature gradients, irrespective of the significant two-mode quantum correlations that may be present in the system. As a measure of such correlations we have analyzed the logarithmic negativity, the tripartite entanglement, characterized by the criteria ${\cal T}_{2,3}$ and ${\cal T}_{3,3}$, and the quantum discord.

Finally, although we have focused on a specific system configuration in which each chain oscillator is connected to an independent heat bath, similar results might be expected for other arrangements, since the emergence of quantum correlations in linear harmonic chains or lattices essentially stems from the proximity of the system to the ground state. The study of the energy current and quantum correlations in systems exhibiting non-linearities deserves further attention.


\section{Summary and concluding remarks}\label{SCon}

We have performed an extensive analysis of the quantum correlations, and the mean value and fluctuations of the energy current across a three-oscillator linear chain at the stationary state, both under Ohmic and super-Ohmic dissipation. We have considered initially squeezed reservoir thermal states, and applied the GLE formalism to determine the correlation functions between the position and momentum operators of the chain oscillators, which completely characterize the stationary properties. 

Interestingly, the results obtained for the quantum correlations are quite similar for both Ohmic and super-Ohmic dissipation. This suggests that the non-Markovian effects do not significantly modify the Markovian results for the {\sl stationary} properties of the quantum correlations in a system of oscillators in contact with independent heat baths. Moreover, the initial squeezing of a heat bath effectively increases the temperature that the  oscillator chain perceives from this bath, and eventually becomes detrimental for the build-up of stationary entanglement.

A different behavior is observed in the case of discord, which can be created by local noisy operations. These quantum correlations highly depend on both the interaction strength between the oscillators and the detuning of their natural frequencies. In particular, the two-mode discord between the side oscillators in the presence of temperature gradients is enhanced at resonance, which indicates that quantum coherence may be favored by thermal non-equilibrium conditions. 

According to our results, both the average and the fluctuations of the stationary energy current across the oscillator chain are mainly determined by the two imposed temperature gradients, and do not seem to be related to the appearance of a rich variety of quantum correlations in the system, comprising two-mode discord and entanglement, bipartite three-mode and genuine tripartite entanglement. The absence of quantum correlation effects in the average energy current can be partially understood in terms of its linear dependence on the correlation terms between the oscillators. In the case of the fluctuations, the more intricate dependence on such terms makes more complex to elucidate their possible connection with quantum correlations.

Nowadays the quantum correlations under thermal non-equilibrium conditions have become a topic of great interest in the fields of quantum information, quantum thermodynamics and the theory of open quantum system. Generally the non-classical correlations, such as entanglement, exhibit a non-linear dependence on the density operator that make difficult to establish any formal connection between them and the response of the system to non-equilibrium constrains. We hope that this work may contribute to stimulate further research in this direction.


\acknowledgments

The authors acknowledge fruitful discussions with L.A. Correa, J.P. Palao and B. Witt. This project was funded by the Spanish MICINN (Grant Nos.\ FIS2010-19998 and FIS2013-41352-P) and by the European Union (FEDER). A.A.V. acknowledges financial support by the Government of the Canary Islands through an ACIISI fellowship (85\% co-financed by the European Social Fund).


\appendix

\section{Susceptibility}\label{ASusc}

In this Appendix we derive the real and imaginary parts of the Fourier transform of the susceptibility $\chi_i(t)$ (\ref{sus}). Considering the spectral density (\ref{Jden}), the imaginary part of the Fourier transform $\tilde{\chi}_i(\omega)$ can be expressed as
\begin{equation}
\text{Im}\left[\,\tilde{\chi}_{i}(\omega)\,\right]=\frac{\pi \hbar}{2}\sum_{\mu=1}^{N}\frac{g_{i\mu}^2}{m_{i\mu} \omega_{i\mu}}\,\left[\,\delta\left(\omega-\omega_{i\mu}\right)-\delta\left(\omega+\omega_{i\mu}\right)\,\right]=\hbar\left[\,\Theta\left(\omega\right)J_{i}\left(\omega\right)-\Theta\left(-\omega\right)J_{i}\left(-\omega\right)\,\right]\,.
\label{APPA1}
\end{equation}
Then the real part can be obtained from the causality of $\chi_i(t)$, according to
\begin{equation}
\text{Re}\left[\,\tilde{\chi}_{i}(\omega)\,\right]=\mathcal{H}\left[\,\text{Im}\left(\chi_{i}(\omega')\right)\,\right](\omega)\doteqdot \frac{1}{\pi}\,P\int\frac{\text{Im}\left[\,\chi_{i}(\omega')\,\right]}{\omega'-\omega}\,d\omega'\,,
\end{equation}
where $\mathcal{H}[\bullet](\omega)$ denotes the Hilbert transform of $\bullet$. Using Eq. (\ref{APPA1}) it follows that
\begin{equation}
\text{Re}\left[\,\chi_{i}(\omega)\,\right]=\mathcal{H}\left[\,\Theta(\omega')J_{i}(\omega')\,\right](\omega)+\mathcal{H}\left[\,\Theta(\omega')J_{i}(\omega')\,\right](-\omega)\,.
\label{APPA2}
\end{equation}
Below are given the expressions of the susceptibility for Ohmic and super-Ohmic spectral densities.


\subsection{Ohmic case}

Assuming the Ohmic spectral density (\ref{SD1}), and considering the well-known properties of the Hilbert transform, the expression (\ref{APPA2}) leads to
\begin{eqnarray}
\text{Re}\left[\,\chi_{i}(\omega)\,\right]&=&\frac{\hbar}{2}\,\pi m \gamma_{i}\left[\,\mathcal{H}\left[\,\Theta(\omega')\,\omega'e^{-\frac{\omega'}{\omega_{c}}}\,\right](\omega)+\mathcal{H}\left[\,\Theta(\omega')\,\omega'e^{-\frac{\omega'}{\omega_{c}}}\,\right](-\omega)\,\right]\nonumber \\
&=&\frac{\hbar}{2}\,\pi m \gamma_{i}\,\omega\left[\,\mathcal{H}\left[ \Theta(\omega')\,e^{-\frac{\omega'}{\omega_{c}}}\,\right](\omega)-\mathcal{H}\left[\,\Theta(\omega')\,e^{-\frac{\omega'}{\omega_{c}}}\,\right] (-\omega)\,\right] +\hbar\,m \gamma_{i}\,\omega_{c} \nonumber \\
&=&\frac{\hbar}{2}\,m \gamma_{i}\,\omega\left[\,e^{-\frac{\omega}{\omega_{c}}}\,\Gamma(0,-\omega/\omega_{c})  -e^{\frac{\omega}{\omega_{c}}}\,\Gamma(0,\omega/\omega_{c})\,\right] +\hbar\,m\gamma_{i}\,\omega_{c}\,,
\end{eqnarray}
where $\Gamma(0,x)=\int^{\infty}_{x}t^{-1}e^{-t}dt$ denotes the incomplete gamma function.


\subsection{Super-Ohmic case}

Following the same procedure as in the previous section for the super-Ohmic spectral density (\ref{SD2}), one obtains
\begin{eqnarray}
\text{Re}\left[\,\chi_{i}(\omega)\,\right]&=&\frac{\hbar}{2\omega_{c}}\,\pi m \gamma_{i}\left[\,\mathcal{H}\left[\,\Theta(\omega')\,(\omega')^2\,e^{-\frac{\omega'}{\omega_{c}}}\,\right](\omega)+\mathcal{H}\left[\,\Theta(\omega')\,(\omega')^2\,e^{-\frac{\omega'}{\omega_{c}}}\,\right] (-\omega)\,\right]\nonumber \\
&=&\frac{\hbar}{2\omega_{c}}\,\pi m\gamma_{i}\,\omega^2\left[\,\mathcal{H}\left[\,\Theta(\omega')\,e^{-\frac{\omega'}{\omega_{c}}}\,\right] (\omega)+\mathcal{H}\left[\,\Theta(\omega')\,e^{-\frac{\omega'}{\omega_{c}}}\,\right](-\omega)\,\right] +\hbar m \gamma_{i}\,\omega_{c} \nonumber \\
&=&\frac{\hbar}{2\omega{c}}\,m \gamma_{i}\,\omega^2\left[\,e^{-\frac{\omega}{\omega_{c}}}\Gamma(0,-\omega/\omega_{c})+e^{\frac{\omega}{\omega_{c}}}\Gamma(0,\omega/\omega_{c})\,\right] +\hbar m \gamma_{i}\,\omega_{c}\,.
\end{eqnarray}


\section{Fluctuating Force}\label{AFF}

In this appendix we derivate the correlation function of the fluctuation forces associated with the heat baths given in Eq.(\ref{CFFDAL}). Considering the time dependence of these forces, it follows 
\begin{eqnarray}
&&\dfrac{1}{2}\left\langle \left\lbrace {\hat F}_{i}(t), {\hat F}_{j}(t')    \right\rbrace  \right\rangle =\dfrac{1}{2}\sum_{\nu,\mu=1}^N g_{i\nu}g_{j\mu} \Bigg( \left\langle \left\lbrace  {\hat x}_{i\nu}(t_{0}), {\hat x}_{j\mu}(t_{0})    \right\rbrace  \right\rangle \cos(\omega_{i\nu}(t-t_{0}))\cos(\omega_{j\mu}(t'-t_{0})) \nonumber \\
&+& \left\langle \left\lbrace  {\hat x}_{i\nu}(t_{0}), {\hat p}_{j\mu}(t_{0})    \right\rbrace  \right\rangle \cos(\omega_{i\nu}(t-t_{0}))\frac{\sin(\omega_{j\mu}(t'-t_{0}))}{m_{j\mu}\omega_{j\mu}}    \nonumber \\
&+&\left\langle \left\lbrace  {\hat p}_{i\nu}(t_{0}), {\hat x}_{j\mu}(t_{0})    \right\rbrace  \right\rangle \cos(\omega_{j\mu}(t'-t_{0}))\frac{\sin(\omega_{i\nu}(t-t_{0}))}{m_{i\nu}\omega_{i\nu}} \nonumber \\
&+&\left\langle \left\lbrace  {\hat p}_{i\nu}(t_{0}), {\hat p}_{j\mu}(t_{0})    \right\rbrace  \right\rangle \frac{\sin(\omega_{i\nu}(t-t_{0}))}{m_{i\nu}\omega_{i\nu}}\frac{\sin(\omega_{j\mu}(t'-t_{0}))}{m_{j\mu}\omega_{j\mu}} \Bigg)\,.
\label{AFF1}
\end{eqnarray}

Replacing the identities (\ref{CMHB}) in Eq.(\ref{AFF1}), and applying the Fourier transform, one obtains
\begin{eqnarray}
&&\frac{1}{2}\int\!\!\int dt\,dt' e^{i\omega t}\,e^{i\omega' t'}\left\langle \left\lbrace {\hat F}_{i}(t), {\hat F}_{i}(t')\right\rbrace\right\rangle =\nonumber \\
&=&\sum_{\mu=1}^N\frac{\hbar g_{i\mu }^2}{m_{i\mu } \omega_{i\mu }}\Big( \left( \frac{1}{2}+N(\omega_{i\mu})\right) \int\!\!\int dt\,dt' e^{i\omega t}\,e^{i\omega' t'}\cos(\omega_{i\mu}(t'-t)) \nonumber \\
&+&\text{Re}\left[M(\omega_{i\mu})\right]  \int\!\!\int dt\, dt' e^{i\omega t}\,e^{i\omega' t'} \cos(\omega_{i\mu}(t+t'-2t_{0}))  \nonumber \\
&+&\text{Im}\left[M(\omega_{i\mu})\right]\int\!\!\int dt\,dt' e^{i\omega t}\,e^{i\omega' t'}\sin(\omega_{i\mu}(t+t'-2t_{0}))\Big)\,.
\nonumber 
\end{eqnarray}
To compute the previous integrals we express the trigonometric functions as complex exponentials, introduce the change of variable $t\rightarrow \tau+t'$ in the first integral, and $t\rightarrow \tau-t'$ in the second and third ones. Secondly, we use the definition of the delta function $2\pi\delta(\omega)=\int dt e^{i\omega t}$ and the identity $1+2N_{th}(\omega_{i\mu})=\coth\left( \frac{\hbar \omega}{2k_{B} T_{i}}\right)$, which lead to
\begin{eqnarray}
&&\frac{1}{2}\left\langle \left\lbrace \tilde{F}_{l}(\omega), \tilde{F}_{l}(\omega')    \right\rbrace  \right\rangle= 2\pi \delta(\omega+\omega')\text{Im}\left[\chi_{l}(\omega)\right]\coth\left( \frac{\hbar \omega}{2k_{B} T_{l}}\right)\cosh(2r_{l}) \nonumber \\
&-&2\pi\hbar\delta(\omega-\omega')\coth\left( \frac{\hbar \omega}{2k_{B} T_{l}}\right)\sinh(2r_{l})\text{Re}\left[e^{i \theta_{l}}\right]\Bigg(\int_{0}^{\infty}d{\bar\omega} {\bar J}_l(\omega-{\bar\omega})e^{2i{\bar\omega} t_{0}} - \int_{0}^{\infty}d{\bar\omega} {\bar J_l}(\omega+{\bar\omega})e^{-2i{\bar\omega} t_{0}} \Bigg)   \nonumber \\
&+& 2i \pi\hbar\delta(\omega-\omega') \coth\left( \frac{\hbar \omega}{2k_{B} T_{l}}\right)\sinh(2r_{l})\text{Im}\left[e^{i \theta_{l}}\right] \Bigg(\int_{0}^{\infty}d{\bar\omega} {\bar J}_l(\omega-{\bar\omega})e^{2i{\bar\omega} t_{0}} + \int_{0}^{\infty}d{\bar\omega} {\bar J}_l(\omega+{\bar\omega})e^{-2i{\bar\omega} t_{0}}\Bigg)\,, \nonumber \\ 
\label{CFFD}
\end{eqnarray}
with $J_l(\omega)=\int d{\bar\omega}{\bar J}_l(\omega-{\bar\omega})$. Before to continue, we pay attention to the four integrals having an explicit dependence on the initial time $t_{0}$. For both the Ohmic and super-Ohmic spectral densities, (\ref{SD1}) and (\ref{SD2}), the corresponding ${\bar J}$ decays more rapidly than $1/{\bar \omega}^2$ at high frequencies. This allows us to use the Riemann-Lebesgue Lemma which states \cite{chandrasekharan19891}
\begin{equation}
\int_{-\infty}^{\infty} f(\omega)e^{i\omega t}d\omega\rightarrow 0 \quad \text{as} \quad t\rightarrow \pm\infty, 
\nonumber
\end{equation}
for $f(\omega)$ an absolutely integrable function in the interval $(-\infty,\infty)$. As a consequence, only the first term in Eq.(\ref{CFFD}) survives after taking the long-time limit $t_{0}\rightarrow -\infty$. The Riemann-Lebesgue Lemma has been successfully employed in the study of the properties of the stationary state for the damped harmonic oscillator \cite{dhar20071,pagel20132}.


\section{Quantum correlations}\label{ASC}
In the following we briefly describe the computation of the logarithmic negativity, quantum discord, and the separability criteria ${\cal T}_{\kappa,n}$.

\subsection{Logarithmic negativity and quantum discord}

The evaluation of the logarithmic negativity (\ref{LogNeg}) requires the symplectic eigenvalues of the partial transpose $\vect V^{T_{j}}$. These are given by the positive square roots of the eigenvalues of $(-i/\hbar) \vect \sigma \vect V_{ij}^{T_{j}} $ \cite{vidal20021}, which are given in terms of the so-called symplectic matrix
\begin{align}\nonumber
\vect\sigma=\begin{bmatrix}
     0 &  \vect{I}_{2}\\
      -\vect{I}_{2}.& 0 \\
\end{bmatrix}\,,
\end{align} 
where $\vect{I}_{n}$ is the $n$-dimensional unit matrix. The corresponding partial transpose matrix is obtained from $ \vect V_{ij}^ {T_{j}}=\vect \Lambda \vect V_{ij} \vect \Lambda $, with 
\begin{align}\nonumber
\vect\Lambda=\vect{I}_{2}\oplus 
\begin{bmatrix}
     \vect{I}_{1} &  0\\
      0 & -\vect{I}_{1} \\
\end{bmatrix} .
\end{align}
 
The evaluation of the quantum discord (\ref{disc}) on the state $\hat\rho_{ij}$ involves an optimization procedure over all positive operator-valued measurements (POVMs) on the $j$-mode, denoted by $\Pi^{(j)}_{l}$. In Eq. (\ref{disc2}), $p_{l}=\text{Tr}_{ij}(\hat \rho_{ij}\Pi^{(j)}_{l})$ is the probability associated with the $lth$ measurement outcome, and $\hat \rho^{(l)}_{i}=\text{Tr}_{j}(\hat\rho_{ij} \Pi^{(j)}_{l})/p_{i}$ is the corresponding post-measurement reduced state of the $i$-mode. An explicit formula providing $D^{\leftarrow}(\hat{\rho}_{ij})$ for any input Gaussian state $\hat{\rho}_{ij}$ is given in \cite{adesso20071}. $D^{\rightarrow}(\hat{\rho}_{ij})$ can be obtained from a similar optimization procedure on the POVMSs in the subsystem $i$. In general, both evaluations of discord may return different values ({\i.e.} $D^{\leftarrow}\neq D^{\rightarrow}$).

\subsection{Separability criteria ${\cal T}_{\kappa,n}$}

Now we describe a hierarchy of separability criteria recently proposed to characterize from genuine multipartite to bipartite entanglement \cite{levi20131}. According to this proposal, the state $\hat \rho$ of a $n$-partite system is $\kappa$-partite entangled, {\sl i.e.} there is at least a entangled subsystem composed of $\kappa$ parties, provide that a given function $\tau_{\kappa,n}(\hat \rho)$ takes a positive value.

The evaluation of the function $\tau_{\kappa,n}$ involves the selection of a set of $2n$ pure states that allows to assess multipartite entanglement. The important point is that a reliable characterization of entanglement requires an appropriate choice of such probe states. However, a priori there is no information about the `optimal' probe states enable to unveil the entanglement  encapsulated by an arbitrary density operator $\hat \rho$. One may circumvent this difficulty by carrying out an optimization procedure over all possible selections in order to obtain the maximum of $\tau_{\kappa,n}$, whose positive value would reveal the entanglement in the system. We denote ${\cal T}_{\kappa,n}$ such maximum.

In  continuous-variable states a Gaussian selection of the probe states provides a readable expression of $\tau_{\kappa,n}$ \cite{valido20141}, which can be optimized with standard procedures, and which is strong enough to detect entanglement for a broad class of Gaussian and non-Gaussian states. In the Gaussian case this expression reads \cite{valido20141},
\begin{eqnarray}
\tau_{\kappa,n}(\hat{\rho})=  \frac{e^{-2\vect X^{T}\vect J_{n}^{T} \frac{1}{\vect \Sigma^{-1} +\vect V^{-1}} \vect J_{n}\vect X }}{\sqrt{\det\left(\vect \Sigma +\vect V \right) }} 
-\sum_{j}a^{(\kappa,n)}_{j}\frac{e^{-\frac{1}{2}\vect X^{T}(\vect P_{j})^{T} \frac{1}{\vect \Sigma +\vect V}\vect P_{j}\vect X }}{\sqrt{\det\left(\vect \Sigma +\vect V \right) }}\ ,
\label{Stau}
\end{eqnarray}
where $a^{(\kappa,n)}_{i}$ are constant values \cite{levi20131}, $\vect X$ is a real $2n$-vector, $\vect J_{n}$ is the standard form of the so-called symplectic matrix, and $\vect P_{j}$ and $\vect \Sigma$ are $2n\times 2n$ real matrices. The objects $\vect X$ and $\vect \Sigma$ denote a  compact form of the first and second moments of the probe states \cite{valido20141}. Hence, the detection of entanglement consists basically in optimizing Eq.(\ref{Stau}) over the variables $\vect X$ and $\vect \Sigma$, {\it i.e.}
\begin{equation}
{\cal T}_{\kappa,n}(\hat{\rho})= \displaystyle\max_{\vect X, \vect \Sigma} \tau_{\kappa,n}(\hat{\rho}).
\nonumber
\end{equation}


\end{document}